\documentclass[prb,twocolumn,showpacs,floatfix]{revtex4}
\usepackage{graphicx}

\catcode`\@=11
\def\simle{\mathrel{\mathpalette\@versim<}}   
\def\simge{\mathrel{\mathpalette\@versim>}}   
\def\@versim#1#2{\lower2.5pt\vbox{\baselineskip0pt \lineskip-.5pt
   \ialign{$\m@th#1\hfil##\hfil$\crcr#2\crcr\sim\crcr}}}
\newcommand{\mib}[1]{\mbox{\boldmath $#1$}} 

\begin{document}

\title{
Orbital and magnetic transitions in geometrically-frustrated vanadium spinels\\
-- Monte Carlo study of an effective spin-orbital-lattice coupled model --
}

\author{
Yukitoshi Motome
}

\affiliation{
RIKEN (The Institute of Physical and Chemical Research), Saitama 351-0198, Japan 
}

\author{
Hirokazu Tsunetsugu
}

\affiliation{
Yukawa Institute for Theoretical Physics, Kyoto University, Kyoto 606-8502, Japan
}

\date{\today}

\begin{abstract}
We present our theoretical and numerical results on 
thermodynamic properties and the microscopic mechanism of two successive transitions 
in vanadium spinel oxides $A$V$_2$O$_4$ ($A$=Zn, Mg, or Cd) 
obtained by Monte Carlo calculations 
of an effective spin-orbital-lattice model 
in the strong correlation limit. 
Geometrical frustration in the pyrochlore lattice structure of V cations 
suppresses development of spin and orbital correlations, however, 
we find that the model exhibits two transitions at low temperatures. 
First, a discontinuous transition occurs with an orbital ordering 
assisted by the tetragonal Jahn-Teller distortion. 
The orbital order reduces the frustration in spin exchange interactions, 
and induces antiferromagnetic correlations in one-dimensional chains 
lying in the perpendicular planes to the tetragonal distortion. 
Secondly, at a lower temperature, a three-dimensional antiferromagnetic order 
sets in continuously, which is stabilized by the third-neighbor interaction 
among the one-dimensional antiferromagnetic chains.
Thermal fluctuations are crucial to stabilize the collinear magnetic state 
by the order-by-disorder mechanism.  
The results well reproduce the experimental data such as 
transition temperatures, 
temperature dependence of the magnetic susceptibility, 
changes of the entropy at the transitions, and 
the magnetic ordering structure at low temperatures. 
Quantum fluctuation effect is also examined
by the linear spin wave theory at zero temperature.
The staggered moment in the ground state is
found to be considerably reduced from saturated
value, and reasonably agrees with the experimental data.
\end{abstract}


\maketitle

\section{Introduction}
\label{Sec:Introduction}

Geometrical frustration in strongly-correlated systems is 
one of the long-standing problems in condensed matter physics. 
Frustration suppresses a formation of a simpleminded long-range order and 
results in nearly-degenerate ground-state manifolds of 
a large number of different states. 
Many well-known examples are found 
in frustrated antiferromagnetic (AF) spin systems. 
There, all the antiparallel spin conditions between interacting pairs 
cannot be satisfied at the same time because closed loops  
contain an odd number of sites. 
The degeneracy due to the frustration yields nontrivial phenomena such as 
complicated ordering structures, spin liquid states, and glassy states. 
\cite{Diep1994,Liebmann1986}
Besides the spin degree of freedom, charge ordering phenomena are also 
much affected by the geometrical frustration. 
\cite{Anderson1956}
Quantum and thermal fluctuations play important roles in these systems, 
which are difficult to handle in a controllable manner. 

\begin{figure}
\caption{
Cubic unit cell of the lattice structure of vanadium spinel oxides 
$A$V$_2$O$_4$. 
(a) The pyrochlore lattice of vanadium cations (red balls). 
(b) The 3D edge-sharing network of VO$_6$ octahedra. 
Oxygen ions are on the corners of octahedra. 
Tetrahedral $A$ sites are omitted. 
} 
\label{fig:lattice structure} 
\end{figure}

Pyrochlore lattice is a typical example of the geometrically-frustrated structures, 
and it consists of a three-dimensional (3D) network of corner-sharing tetrahedra
as shown in Fig.~\ref{fig:lattice structure} (a). 
Spin systems on the pyrochlore lattice have been intensively studied. 
\cite{Liebmann1986,Canals2000,Harris1991,Tsunetsugu2001,Koga2001,Berg2003,Reimers1991,Moessner1998}
In particular, for quantum $S=1/2$ spin systems 
with only nearest-neighbor interactions, it is predicted that 
a macroscopic number of singlet states lie inside the singlet-triplet gap.
\cite{Canals2000}
Several types of symmetry breakings are predicted within the singlet subspace, 
e.g., dimer/tetramer ordering, but without magnetic long-range ordering. 
\cite{Harris1991,Tsunetsugu2001,Koga2001,Berg2003}
Antiferromagnetic (AF) classical spin systems on the pyrochlore lattice are also believed 
to show no long-range ordering at any temperature. 
\cite{Liebmann1986,Reimers1991,Moessner1998}
Due to the three dimensionality and 
the large unit cell (16 sites in the cubic unit cell), 
pyrochlore systems remain a big challenge to theoreticians and 
are still far from comprehensive understanding. 

We can find many pyrochlore systems in real compounds. 
Most typically, pyrochlore systems are realized in so-called $B$ spinel oxides, 
where only $B$-site cations are magnetic in the general chemical formula 
$AB_2$O$_4$. 
Figure \ref{fig:lattice structure} (b) shows the $B$ spinel structure, 
which consists of a network of edge-sharing $B$O$_6$ octahedra. 
$B$ cation is shown by ball in center of each octahedron, and 
octahedron corners are occupied by oxygen ions, while 
nonmagnetic $A$-site cations are not shown for simplicity. 
With omitting oxygen ions on the corners of the octahedra, 
one obtains the pyrochlore lattice of $B$ cations 
in Fig.~\ref{fig:lattice structure} (a). 

In this paper, we will investigate insulating vanadium spinel oxides, 
$A$V$_2$O$_4$ with divalent $A$-site cations such as Zn, Mg, or Cd. 
In these compounds, each V$^{3+}$ cation has two $3d$ electrons 
in a high-spin state by the Hund's-rule coupling, and 
they are Mott insulators. 
\cite{Muhtar1988} 
Thus, we may consider that pyrochlore spin systems with $S=1$ are realized. 
Curie-Weiss temperature is estimated from the magnetic susceptibility 
as $|\Theta_{\rm CW}| \sim 1000$K, and 
any long-range ordering does not occur down to significantly 
lower temperatures than $|\Theta_{\rm CW}|$ 
due to the geometrical frustration. 
\cite{Muhtar1988} 
For instance, in ZnV$_2$O$_4$, a structural phase transition occurs 
at $T_{\rm c1} \simeq 50$K 
from the high-temperature cubic phase to the low-temperature tetragonal phase 
with a flattening of VO$_6$ octahedra in the $c$ direction. 
\cite{Ueda1997}
Successively, an AF transition occurs at $T_{\rm c2} \simeq 40$K. 
\cite{Ueda1997}
Neutron scattering experiments revealed that 
the AF ordering structure below $T_{\rm c2}$ is a collinear one 
which consists of the staggered AF chains in the $ab$ planes stacking 
in the $c$ direction with a four-times period 
as up-up-down-down-$\cdot\cdot\cdot$ 
as shown in Fig.~\ref{fig:spin order}. 
\cite{Niziol1973,Izyumov1979} 
These two successive transitions are commonly seen in compounds 
MgV$_2$O$_4$ and CdV$_2$O$_4$. 
\cite{Mamiya1997,Nishiguchi2002} 
This indicates that the degenerate ground-state manifolds 
in the pyrochlore systems 
are lifted at low temperatures in some manner. 

\begin{figure}
\caption{
Spin ordering structure proposed for ZnV$_2$O$_4$ 
on the basis of the neutron scattering results. 
The ordering pattern consists of staggered AF chains 
in the $ab$ plane (red solid lines) which stack 
with a four-times period in the $c$ direction 
as up-up-down-down-$\cdot\cdot\cdot$ (blue dashed lines). 
(b) Projection of (a) from the $z$ direction. 
Symbols of $+$ and $-$ denote the up and down spins, respectively. 
} 
\label{fig:spin order} 
\end{figure}

A few years ago, Yamashita and Ueda proposed a scenario to explain 
the mechanism of the transitions in $A$V$_2$O$_4$. 
\cite{Yamashita2000}
Their approach is based on a valence-bond-solid picture for $S=1$ spins
and takes account of the coupling to Jahn-Teller (JT) lattice distortions. 
They claimed that the first transition at $T_{\rm c1}$ is 
due to the JT effect which lifts the degeneracy of 
the spin-singlet local ground states at each tetrahedron unit. 
This scenario based on the spin-JT coupling is appealing, 
however, some difficulty still remains. 
The problem is that it is difficult to explain the magnetic transition 
at a lower temperature $T_{\rm c2}$. 
In this approach, a finite energy gap is assumed
between the spin-singlet ground-state subspace and 
the spin-triplet excitations, and 
a low-energy effective theory is derived to describe 
a phase transition within the spin-singlet subspace. 
High-energy excitations with total spin $S \neq 0$ are already 
traced out at the starting point, and therefore 
their effective model has 
no chance to describe the AF ordering within their theory. 

A similar JT scenario was also examined for classical spin systems. 
\cite{Tchernyshyov2002} 
In this case, although the problem to have AF order does not exist, 
there is another difficulty to explain the following generic difference 
from chromium family of $A$Cr$_2$O$_4$ ($A$=Zn, Mg, or Cd). 
These chromium oxides are also $B$ spinels and magnetic Cr cations constitute 
a pyrochlore lattice. 
However, in contrast to the two transitions in vanadium compounds, 
the chromium compounds exhibit only one transition, 
i.e., the AF order appears simultaneously with the structural transition. 
\cite{Kagomiya2002,Rovers2002} 
This clear difference is generic, being independent of divalent $A$ cations, and 
ascribed to the difference of magnetic cations V$^{3+}$ and Cr$^{3+}$, 
which cannot be explained by the classical spin approach based on the spin-JT effect 
unless there exists essential difference in the model parameters 
between the two families. 

Therefore, these spin-JT type theories appear to be insufficient 
to explain the mechanism of two transitions in vanadium spinels 
$A$V$_2$O$_4$. 
These insulating compounds are undoped states of LiV$_2$O$_4$ 
which exhibits a unique heavy fermion behavior. 
\cite{Kondo1997,Urano2000}
The origin of the mass enhancement is still controversial 
between the scenario based on the Kondo effect 
\cite{Anisimov1999,Singh1999,Nekrasov2003}
and the scenario of strong correlations with the geometrical frustration. 
\cite{Eyert1999,Matsuno1999,Fujimoto2001,Tsunetsugu2002,Yamashita2003}
Since the doping of Li shows systematic changes of magnetic
\cite{Muhtar1988} 
and transport properties,
\cite{Kawakami1986} 
as well as the phase diagram, 
\cite{Ueda1997,Onoda1997}
understanding of undoped materials 
may give a starting point to discuss the doped state in LiV$_2$O$_4$. 
Therefore, it is also highly desired to clarify the mechanism of 
orderings in the undoped compounds $A$V$_2$O$_4$. 

The generic difference between vanadium and chromium spinels mentioned above 
suggests an importance of $t_{2g}$ orbital degrees of freedom. 
In the case of chromium spinels, 
each Cr$^{3+}$ cation has three electrons in threefold $t_{2g}$ levels 
and large Hund's-rule coupling leads to a high-spin state, and 
therefore, there is no orbital degree of freedom. 
On the contrary, in the case of vanadium spinels, 
since each V$^{3+}$ cation has two electrons, 
the orbital degree of freedom is active. 
With taking account of this $t_{2g}$ orbital degeneracy, 
the authors have derived an effective spin-orbital-lattice coupled model 
in the strong correlation limit 
and investigated it by mean-field type arguments. 
\cite{Tsunetsugu2003}
A reasonable scenario was obtained, but 
discussions were limited to a qualitative level. 
In order to investigate temperature dependences of physical properties 
semiquantitatively accurate enough to be compared 
with experimental data, 
we need more elaborate analysis. 

In the present study, we will investigate thermodynamic properties of 
the effective spin-orbital-lattice model derived by the authors 
in Ref.~\onlinecite{Tsunetsugu2003} by extensive Monte Carlo (MC) calculations. 
We will show that this model indeed exhibits two successive transitions 
in a reasonable parameter range, and 
clarify the microscopic mechanism of these transitions in detail.  
First, an orbital order appears 
with the tetragonal JT distortion which flattens VO$_6$ octahedra. 
This orbital order reduces magnetic frustration partially, and 
enhances AF spin correlations in one-dimensional (1D) chains 
in the $ab$ planes. 
At a lower temperature, the third-neighbor exchange interaction and 
thermal fluctuations align these 1D AF chains coherently, and 
stabilize a 3D collinear AF order. 
With comparing numerical results of physical quantities with experimental data, 
we will show that our theory captures essential physics at low temperatures 
in the vanadium spinels $A$V$_2$O$_4$. 

This paper is organized as follows.
In Sec.~\ref{Sec:Model and Method}, we introduce the effective 
spin-orbital-lattice coupled model, and briefly summarize 
the mean-field arguments discussed in Ref.~\onlinecite{Tsunetsugu2003}. 
Realistic parameter values and MC method are also described. 
In Sec.~\ref{Sec:Results}, we show numerical results 
in comparison with experimental data. 
We then make several remarks, in particular, 
on comparisons with experimental results and 
other theoretical proposals in Sec.~\ref{Sec:Discussions}. 
Section \ref{Sec:Summary} is devoted to summary.

\section{Model and Method}
\label{Sec:Model and Method}

\subsection{Effective spin-orbital-lattice model}
\label{Sec:Spin-Orbital Model}

In the present study, we will investigate thermodynamic properties of 
the spin-orbital-lattice coupled model which is proposed 
by the authors in Ref.~\onlinecite{Tsunetsugu2003}. 
The Hamiltonian consists of two terms as
\begin{equation}
H = H_{\rm SO} + H_{\rm JT}.
\label{eq:H}
\end{equation}
The first term describes exchange interactions in spin and orbital 
degrees of freedom and the second term is for orbital-lattice couplings 
of the Jahn-Teller type.  

The spin-orbital Hamiltonian $H_{\rm SO}$ 
is derived from a multiorbital Hubbard model 
with threefold $t_{2g}$ orbital degeneracy 
by the perturbation in the strong correlation limit. 
\cite{Kugel1973}
The starting $t_{2g}$ Hubbard model is given in the form  
\begin{eqnarray}
&& H_{\rm Hub} = \sum_{i,j} \sum_{\alpha,\beta} \sum_{\tau}
[ t_{\alpha\beta} (\mib{r}_i - \mib{r}_j) 
c_{i \alpha \tau}^{\dagger} c_{j \beta \tau} + {\rm H.c.} ]
\nonumber
\\
&& \quad + \frac12 \sum_i \sum_{\alpha\beta,\alpha'\beta'} \sum_{\tau\tau'} 
U_{\alpha\beta,\alpha'\beta'}
c_{i\alpha\tau}^{\dagger} c_{i\beta\tau'}^{\dagger}
c_{i\beta'\tau'} c_{i\alpha'\tau},
\label{eq:H_multiorbital}
\end{eqnarray}
where $i,j$ and $\tau,\tau'$ are site and spin indices, respectively, 
and $\alpha,\beta = 1$ ($d_{yz}$), $2$ ($d_{zx}$), $3$ ($d_{xy}$) are orbital indices. 
The first term of $H_{\rm Hub}$ is the electron hopping, and 
the second term describes Coulomb interactions, 
for which we use the standard parametrizations,
\cite{note2}
\begin{eqnarray}
U_{\alpha\beta,\alpha'\beta'} &=& U' \delta_{\alpha\alpha'} \delta_{\beta\beta'} 
+ J_{\rm H} (\delta_{\alpha\beta'} \delta_{\beta\alpha'} +
\delta_{\alpha\beta} \delta_{\alpha'\beta'}), 
\\
U &=& U' + 2J_{\rm H}.
\end{eqnarray}
We do not include here the relativistic spin-orbit coupling and 
the trigonal distortion in the model (\ref{eq:H_multiorbital}). 
Effects of these neglected elements will be discussed 
in Sec.~\ref{Sec:symmetry} and E. 

Considering the vanadium spinel oxides are insulators, 
it is reasonable to start from the strong correlation limit 
and treat the hopping term as perturbation. 
The unperturbed states are atomic eigenstates with 
two electrons on each V cation in a high-spin state. 
As for the perturbation part, 
on the basis of the tight-binding fit for the band structure
\cite{Matsuno1999} 
(see Sec.~\ref{Sec:Parameters}),
we take account of hopping integrals of $\sigma$ bonds
for nearest-neighbor pairs, $t_{\sigma}^{\rm nn}$, and 
for third-neighbor pairs, $t_{\sigma}^{\rm 3rd}$, 
as shown in Fig.~\ref{fig:transfer}. 
Note that the third-neighbor pair corresponds to 
the next nearest-neighbor pair along each chain. 
The hopping integral between second-neighbor pairs 
(the gray arrow in Fig.~\ref{fig:transfer}) is expected to be small 
because of the geometry of the pyrochlore lattice, 
\cite{note4}
and moreover the exchange interaction derived from it is frustrated. 

\begin{figure}
\caption{
Hopping integrals for the $\sigma$ bonds; 
$t_{\sigma}^{\rm nn}$ and $t_{\sigma}^{\rm 3rd}$ are 
for the nearest-neighbor sites and for the third-neighbor sites, 
respectively. 
The overlaps between $d_{xy}$ orbitals within the $xy$ plane 
(the solid arrows) and 
those between $d_{yz}$ orbitals within the $yz$ plane 
(the dashed arrows) are shown. 
The $d_{zx}$ overlaps are similarly taken into account. 
The gray arrow shows an example of second-neighbor pair. 
} 
\label{fig:transfer} 
\end{figure}

The second order perturbation in $t_{\sigma}^{\rm nn}$ and 
$t_{\sigma}^{\rm 3rd}$ gives the Hamiltonian in the form
\begin{eqnarray}
&& H_{\rm SO} = H_{\rm SO}^{\rm nn} + H_{\rm SO}^{\rm 3rd}, 
\label{eq:H_SO}
\\
&& H_{\rm SO}^{\rm nn} = -J \sum_{\langle i,j \rangle} \
[ \ h_{\rm o-AF}^{(ij)} + h_{\rm o-F}^{(ij)} \ ],
\label{eq:H_nn}
\\
&& H_{\rm SO}^{\rm 3rd} = -J_3 \sum_{\langle\!\langle i,j \rangle\!\rangle} \
[ \ h_{\rm o-AF}^{(ij)} + h_{\rm o-F}^{(ij)} \ ],
\label{eq:H_3rd}
\\
&& h_{\rm o-AF}^{(ij)} = (A + B \mib{S}_i \cdot \mib{S}_j)
\nonumber \\
&& \quad \times
\big[ n_{i \alpha(ij)} (1 - n_{j \alpha(ij)}) 
+ (1 - n_{i \alpha(ij)}) n_{j \alpha(ij)} \big],
\label{eq:h_o-AF}
\\
&& h_{\rm o-F}^{(ij)} = C (1 - \mib{S}_i \cdot \mib{S}_j)
n_{i \alpha(ij)} n_{j \alpha(ij)}, 
\label{eq:h_o-F}
\end{eqnarray}
where $\mib{S}_i$ is the $S=1$ spin operator 
and $n_{i \alpha} = \sum_{\tau} c_{i \alpha \tau}^\dagger c_{i \alpha \tau}$ 
is the density operator for site $i$ and orbital $\alpha$. 
The summations with $\langle i,j \rangle$ and 
$\langle\!\langle i,j \rangle\!\rangle$ 
are taken over the nearest-neighbor sites and third-neighbor sites, respectively. 
Here, $\alpha(ij)$ is the orbital which has a finite hopping integral
between the sites $i$ and $j$, for instance, 
$\alpha(ij) = 3$ ($d_{xy}$) for $i$ and $j$ sites in the same $xy$ plane. 
The other parameters in Eqs.~(\ref{eq:H_nn})-(\ref{eq:h_o-F}) are determined  
by coupling constants in Eq.~(\ref{eq:H_multiorbital}) as 
\cite{note}
\begin{eqnarray}
&& J = (t_{\sigma}^{\rm nn})^2 / U,
\label{eq:J}
\\
&& J_3 = (t_{\sigma}^{\rm 3rd})^2 / U,
\label{eq:J_3}
\\
&& A = (1-\eta)/(1-3\eta),
\label{eq:A}
\\
&& B = \eta/(1-3\eta),
\label{eq:B}
\\
&& C = (1+\eta)/(1+2\eta),
\\
&& \eta = J_{\rm H} / U,
\label{eq:eta}
\end{eqnarray}
and each site is subject to the local constraint, 
$\sum_{\alpha=1}^3 n_{i \alpha} = 2$. 
Realistic values of these parameters are given in Sec.~\ref{Sec:Parameters}. 

An important feature of $H_{\rm SO}$ is 
the highly anisotropic form of the orbital intersite interaction. 
It is a three-state clock type interaction 
corresponding to three different orbital states, 
in which there is no quantum fluctuation 
since the density operator $n_{i \alpha}$ is a constant of motion.  
This anisotropy comes from the orbital diagonal nature of 
the $\sigma$-bond hopping integrals 
which do not mix different orbitals. 
Moreover, the orbital interaction depends on 
both the bond direction and the orbital states in two sites. 
On the other hand, the spin exchange interaction is Heisenberg type 
and isotropic, independent of the bond direction. 

\begin{figure}
\caption{
(a) Tetragonal distortion and the level splitting. 
(b) The orbital ordering pattern for the model (\ref{eq:H}) 
predicted by the mean-field argument in Ref.~\onlinecite{Tsunetsugu2003}. 
The ferro-type (antiferro-type) orbital bonds are shown 
by the blue solid (red dashed) lines. 
} 
\label{fig:orbital order} 
\end{figure}

The orbital-lattice term $H_{\rm JT}$ in Eq.~(\ref{eq:H}) reads 
\begin{eqnarray}
H_{\rm JT} &=& \gamma \sum_i Q_{i} (n_{i1} + n_{i2} - 2 n_{i3})
\nonumber \\
&+& \sum_i Q_i^2 / 2 
- \lambda \sum_{\langle i,j \rangle} Q_i Q_j, 
\label{eq:H_JT}
\end{eqnarray}
where $\gamma$ is the electron-phonon coupling constant and 
$Q_i$ denotes the amplitude of local lattice distortion at site $i$. 
Here, we take account of only the tetragonal mode in the $z$ direction. 
Note that this simplification breaks the cubic symmetry of the system. 
We choose the sign of $Q_i$ such that 
it is positive for a flattening of VO$_6$ octahedra 
which leads to the level splitting in Fig.~\ref{fig:orbital order} (a).  
The second term in Eq.~(\ref{eq:H_JT}) 
denotes the local elastic energy of distortions. 
Finally, the third term denotes 
the interaction of JT distortions between nearest-neighbor sites 
which mimics the cooperative aspect of the JT distortion. 
It is reasonable to assume a positive value of $\lambda$ 
because a tetragonal distortion of a VO$_6$ octahedron modifies 
its neighboring octahedra in a similar distortion 
due to the edge-sharing 3D network of octahedra 
in Fig.~\ref{fig:lattice structure} (b). 
For simplicity, we here neglect quantum nature of phonons. 
Although JT distortions modify $H_{\rm SO}$ through changes of 
hopping integrals, we neglect these corrections in the present study. 

We normalized the variable $Q_i$ 
to absorb the elastic constant in the second term in Eq.~(\ref{eq:H_JT}), 
hence the dimension of $Q_i$ is (energy)$^{1/2}$. 
As a result, the dimensions of $\gamma$ and $\lambda$ are 
(energy)$^{1/2}$ and (energy), respectively. 

For the following discussions, 
we here briefly summarize the results of the mean-field type analysis 
on the model (\ref{eq:H}) obtained in Ref.~\onlinecite{Tsunetsugu2003}. 
The analysis predicts that first, the degeneracy due to the geometrical 
frustration will be partially lifted in the orbital channel. 
There, the anisotropy of the orbital interaction 
in $H_{\rm SO}^{\rm nn}$ plays a crucial role; 
the interaction is a three-state clock type and 
depends on the orbital states as well as the bond direction. 
The remaining degeneracy of orbital states 
is lifted by the tetragonal JT coupling in $H_{\rm JT}$:
A flattening of VO$_6$ octahedra splits the threefold orbital levels 
as shown in Fig.~\ref{fig:orbital order} (a), and
selects the orbital ordering structure 
as shown in Fig.~\ref{fig:orbital order} (b). 
There, one of the two electrons occupies the $d_{xy}$ orbital at every site, and 
the other occupies either $d_{yz}$ or $d_{zx}$ orbital 
in an alternative manner along the $z$ direction. 
When we consider only the nearest-neighbor interactions $H_{\rm SO}^{\rm nn}$, 
this orbital occupation 
induces AF spin interactions on the bonds within the $ab$ planes 
[the blue solid lines in Fig.~\ref{fig:orbital order} (b)] and 
ferromagnetic spin interactions on the bonds 
among the $ab$ planes 
[the red dashed lines in Fig.~\ref{fig:orbital order} (b)]. 
The coupling constant for the former AF interaction is $JC$ and 
that for the latter ferromagnetic interaction is $-JB$. 
Since $\eta$ is a small parameter of the order of 0.1 
as estimated in Sec.~\ref{Sec:Parameters}, 
the former AF interaction is much larger than the latter ferromagnetic one. 
Moreover, the ferromagnetic interactions are frustrated 
for the AF spin configuration within the $ab$ planes 
because of the geometry of the pyrochlore lattice. 
Hence, under the orbital ordering shown in Fig.~\ref{fig:orbital order} (b), 
the AF spin correlations develop within the 1D chains in the $ab$ planes, and 
the 1D AF chains are independent with each other; 
relative angles among the AF moments are not yet determined 
at this stage. 
The relative angles are partially fixed 
by including the third-neighbor interactions $H_{\rm SO}^{\rm 3rd}$: 
The third-neighbor interactions align the AF moments 
in two next-neighboring $ab$ planes, and 
lead to a 3D collinear AF order with the wave vector $\mib{q} = (0,0,2\pi/c)$. 
(See Fig.~\ref{fig:two sublattices}.) 
In the ordered state, however, 
there are two independent AF sublattices;
one consists of $[110]$ chains and the other consists of $[1\bar1 0]$ chains. 
The relative angle between the AF moments on the two sublattices 
is still free in this mean-field argument. 
From the spin wave calculation of the zero-point energy, 
we discussed that quantum fluctuations 
fix the relative angle and 
stabilize a collinear AF order in Fig.~\ref{fig:spin order} 
that is consistent with the neutron scattering result. 

The mean-field argument gives a reasonable scenario for two transitions 
in $A$V$_2$O$_4$, but the argument is limited to a qualitative level. 
In order to confirm the scenario and understand the experimental results 
more quantitatively, we need more sophisticated analysis, 
especially for the thermodynamic properties of the system. 
In the present study, we will perform 
the Monte Carlo simulation for this purpose.

\subsection{Parameters}
\label{Sec:Parameters}

Here, we estimate realistic values of parameters 
in the model (\ref{eq:H}) which are given by the parameters 
in the starting $t_{2g}$ Hubbard model (\ref{eq:H_multiorbital}). 
As for the hopping parameters $t_{\alpha\beta} (\mib{r}_i - \mib{r}_j)$ 
in $H_{\rm Hub}$, 
a tight-binding fit to the results of the first-principle band calculation 
suggests the dominant hopping integrals are those of $\sigma$ bonds, 
and gives  
$t_{\sigma}^{\rm nn} \sim -0.32$eV and 
$t_{\sigma}^{\rm 3rd} \sim -0.045$eV 
for nearest-neighbor and third-neighbor pair of sites, respectively. 
\cite{Matsuno1999,note4} 
For Coulomb interactions, there are estimates 
based on the cluster analysis for optical experiments 
for vanadium perovskites $A$VO$_3$, which also have a VO$_6$ octahedral unit.  
\cite{Mizokawa1996}
The estimates are $U \sim 6$eV and $J_{\rm H} \sim 0.68$eV, 
thus, $\eta = J_{\rm H}/U$ in Eq.~(\ref{eq:eta}) 
is a small parameter of the order of 0.1. 
We will set $\eta=0.08$ in the following numerical calculations. 
The estimates of $t_{\sigma}^{\rm nn}$, $t_{\sigma}^{\rm 3rd}$, and $U$ 
give $J \sim 200$K and $J_3 \sim 4$K, i.e., $J_3/J \sim 0.02$. 
In the Monte Carlo calculations, 
we study mainly the case of $J_3/J = 0.02$, 
but we vary the value of $J_3/J$ from 0 to 0.05 
to examine the systematic change by $J_3$. 
Hereafter, we will set $J=1$ as an energy unit and 
the lattice constant of cubic unit cell as a length unit ($a=b=c=1$), and 
use the convention of the Boltzmann constant $k_{\rm B} = 1$. 

It is hard to estimate 
the electron-phonon interaction parameters $\gamma$ and $\lambda$, 
and therefore, we treat them as variable parameters in the present study.  
In the present MC calculations to confirm the above mean-field scenario, 
we are interested in the parameter region where 
the orbital order in Fig.~\ref{fig:orbital order} (b) is stabilized 
by the tetragonal JT distortion with a flattening of VO$_6$ octahedra. 
The stability conditions for this orbital and lattice order will be obtained 
in Appendix A by a mean-field type argument.
In the following MC calculations, 
we will show MC results for typical values of $\gamma$ and $\lambda$ 
which satisfy the conditions as $\gamma^2/J = 0.04$ and $\lambda/J = 0.15$.

\subsection{Monte Carlo method}
\label{Sec:Monte Carlo}

In the present study, we will use MC calculations 
to investigate thermodynamic properties of 
the effective spin-orbital-lattice model (\ref{eq:H}). 
Quantum MC simulations for frustrated systems are known 
to be difficult because of the negative sign problem. 
In the present MC study, 
we neglect quantum fluctuations and approximate the model in the classical level. 
This approximation retains effects of thermal fluctuations 
which may play dominant roles in finite-temperature transitions. 
The quantum nature originates only from spin $S=1$ operators 
in the model (\ref{eq:H}), 
since the orbital interaction is classical and 
is a diagonal one of three-state clock type and 
since JT distortions are also treated as classical variables. 
Thus, we approximate the spin operators by classical vectors 
with the modulus $|\mib{S}| = 1$ 
[the length of the vector is normalized 
to give the same largest $z$ component $S^z=1$ (classical part)]. 
Effects of quantum fluctuations will be discussed by using 
the spin wave approximation in Sec.~\ref{Sec:quantum fluctuation}. 
Thereby, the model (\ref{eq:H}) consists of the classical Heisenberg part for spins, 
the three-state clock part for orbitals, and the classical phonon part. 
We use a standard metropolis MC algorithm. 

In the actual MC calculations, the MC sampling is performed to measure 
spin vectors $\mib{S}_i$, 
three-state clock spins for the orbital states
(defined in Sec.~\ref{Sec:Orbital Ordering}), and 
amplitudes of the JT distortion $Q_i$ at all the lattice sites. 
We typically perform $10^5$ MC samplings for measurements 
after $10^5$ steps for thermalization. 
The measurements are performed in every $N_{\rm int}$-times MC update, 
and we typically take $N_{\rm int} = 2$. 
Results are divided into five bins to estimate statistical errors 
by variance of average values in the bins. 
Here, one MC update consists of several-times sweeps (typically twice) 
for spin directions, orbital states, and JT distortions. 
The one sweep is $N_{\rm site}$-times trials 
by choosing a site randomly, 
where $N_{\rm site}$ is the number of lattice sites. 
In the sampling on spin directions, we apply the so-called pivot rotation 
when a trial is rejected, which is a precession without energy cost: 
The pivot rotation of a spin is achieved by a random rotation 
with keeping the relative angle to the mean-field vector 
determined by its nearest-neighbor and third-neighbor spin and orbital states. 
This accelerates the MC sampling in the configurational space. 

As shown in Sec.~\ref{Sec:Results}, 
the orbital transition accompanied by the JT distortion is first order. 
To avoid a hysteresis and determine the transition temperature precisely, 
we start MC calculations at each temperature 
from a mixed initial condition for orbital and lattice states in which 
a half of the system takes a low-temperature ordered configuration and 
the rest takes a high-temperature disordered configuration. 
This technique is known to be free from trapping 
at a metastable state for large enough system sizes. 
\cite{Ozeki2003}
At very low temperatures, we use an perfectly ordered initial state 
to accelerate the convergence. 
The system sizes in the present work are up to $L = 12$ 
where $L$ is the linear dimension of the system 
measured in the cubic units, i.e., 
the total number of sites $N_{\rm site}$ is given by 
$L^3 \times 16$.

\section{Results}
\label{Sec:Results}

In this section, we present MC results for the model (\ref{eq:H}) 
in comparison with experimental data. 
In Secs.~\ref{Sec:Transitions}-C, 
we show the results for the typical case of $J_3/J = 0.02$. 
Systematic changes with $J_3/J$ and a generic phase diagram 
will be discussed in Sec.~\ref{Sec:Phase Diagram}. 

\subsection{Two transitions}
\label{Sec:Transitions}

In the following, we present MC results for $J_3/J = 0.02$ 
to show that the model (\ref{eq:H}) exhibits two phase transitions with temperature. 
In Sec.~\ref{Sec:E and C}, we present the MC results for 
the internal energy and the specific heat, which 
show two different anomalies. 
We also discuss the changes of the entropy related to the two transitions. 
In Secs.~\ref{Sec:Orbital Ordering} and 3, 
we discuss the nature of the two transitions 
by calculating the order parameters. 
In Sec.~\ref{Sec:Collinearity}, we examine the magnetic ordering structure 
in the low temperature phase, and point out the importance of thermal fluctuations. 
Sec.~\ref{Sec:chi} contains MC results for the uniform magnetic susceptibility 
for comparison with experimental data. 

\subsubsection{Internal energy and specific heat}
\label{Sec:E and C}

\begin{figure}
\includegraphics[width=7cm]{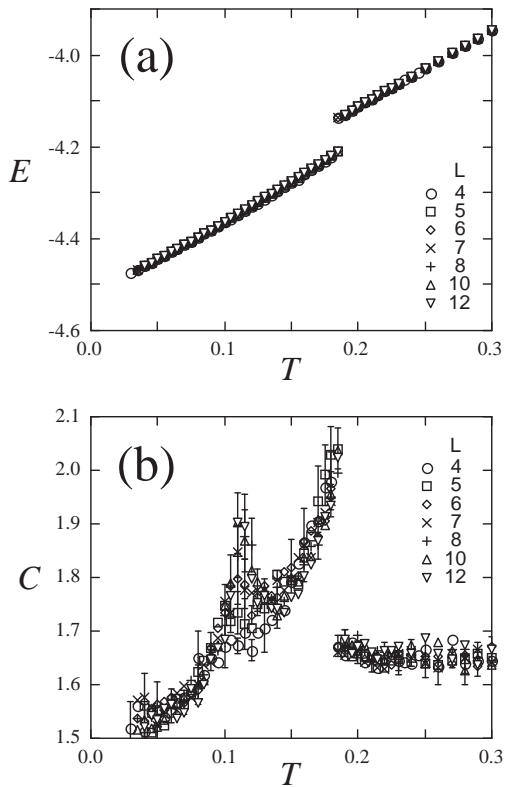}
\caption{
(a) The internal energy per site [Eq.~(\ref{eq:E})] and 
(b) the specific heat per site [Eq.~(\ref{eq:C})] 
at $J_3/J = 0.02$.
Error bars are smaller than symbol sizes in (a). 
Typical error bars are shown in (b). 
} 
\label{fig:E and C} 
\end{figure}

Figure~\ref{fig:E and C} shows temperature dependences of 
the internal energy and the specific heat per site. 
The internal energy per site is calculated by the thermal average of 
the Hamiltonian (\ref{eq:H}) as 
\begin{equation}
E = \langle H \rangle / N_{\rm site}.
\label{eq:E} 
\end{equation} 
The specific heat is calculated by fluctuations of the internal energy as 
\begin{equation}
C = \frac{\langle H^2 \rangle - \langle H \rangle^2}{T^2 N_{\rm site}}. 
\label{eq:C} 
\end{equation}

As shown in Fig.~\ref{fig:E and C} (a), the internal energy $E$ 
jumps at $T \simeq 0.19J$. 
It indicates that a first-order transition occurs at this temperature. 
The jump is also found in the specific heat at the same temperature 
in Fig.~\ref{fig:E and C} (b). 
The specific heat shows another anomaly at a lower temperature $T \simeq 0.115J$. 
There, we find a systematic enhancement of the peak as the system size increases, 
which is a sign of second-order phase transition. 
Thus, MC data in Fig.~\ref{fig:E and C} indicate that 
the system shows two different transitions; 
the first-order transition at $T_{\rm O} \simeq 0.19J$ and 
the second-order transition at $T_{\rm N} \simeq 0.115J$. 
In the following sections, the two transitions are to be assigned to 
the orbital ordering with tetragonal lattice distortion and 
the AF spin ordering, respectively. 

It is noted that 
the specific heat approaches $3/2$ (in units of $k_{\rm B}$) 
as $T \rightarrow 0$ as shown in Fig.~\ref{fig:E and C} (b). 
A finite value of $C$ at $T=0$ is characteristic to classical models, and 
one degree of freedom remaining at the ground state 
contributes $1/2$ to $C$. 
Hence, the data in Fig.~\ref{fig:E and C} (b) suggests that 
there remain three degrees of freedom at $T=0$ in the present model. 
From the discussions of the ordered phase at low temperatures 
in the following sections, they can be ascribed to 
two transverse modes of the AF spin order and one JT mode. 

From the jump of the internal energy in Fig.~\ref{fig:E and C} (a), 
we estimate the entropy jump $\Delta {\cal S}$ 
associated with the first-order transition 
from the disordered phase above $T_{\rm O}$ 
to the ordered phase below $T_{\rm O}$. 
The two phases have the same free energy $F = E - T{\cal S}$ at $T_{\rm O}$, 
which gives us the entropy difference $\Delta {\cal S}$ as 
\begin{equation}
\Delta {\cal S} = \lim_{\delta \rightarrow 0} 
\frac{E(T_{\rm O} + \delta) - E(T_{\rm O} - \delta)}{T_{\rm O}} 
\sim 0.4. 
\end{equation}
This corresponds to $\sim 30$-$40$\% of $\ln 3$ per site.  

The amount of the entropy which is related to 
the fluctuation around the second-order transition at $T_{\rm N}$ 
is estimated from the area of the anomalous peak at $T_{\rm N}$ 
in the plot of $C/T$ as a function of $T$. 
Although it is difficult to estimate it 
because of the large system-size dependence of $C$ as well as 
the large error bars in Fig.~\ref{fig:E and C} (b), 
a rough estimate is obtained by an interpolation 
with polynomial functions for the normal contribution and 
a numerical integration of the anomalous part. 
The estimate is roughly $0.05$-$0.1$, which corresponds to 
$\sim 5$-$10$\% of $\ln 3$ per site. 

In experiments, the amounts of entropy related to the two transitions 
are also estimated from the specific heat. 
In ZnV$_2$O$_4$, the entropy change in the higher-temperature transition 
at $T_{\rm c1}$ is $\sim 3$-$4$J/mol~K, i.e., $\sim 20$\% of $\ln 3$ per V cation. 
\cite{Kondo2000}
The entropy related to the lower-temperature transition at $T_{\rm c2}$ 
is small and not estimated quantitatively, 
but roughly less than $1$J/mol~K, i.e., 
$\simle 5$\% of $\ln 3$ per V cation. 
\cite{Kondo2000}
In MgV$_2$O$_4$, the former is estimated as $\sim 3$J/mol~K, i.e., 
$\sim 16$\% of $\ln 3$, and
the latter is $\sim 0.4$J/mol~K, i.e., $\sim 2$\% of $\ln 3$ per V cation. 
\cite{Mamiya1997}
Our estimates from the MC results in Fig.~\ref{fig:E and C} 
show semiquantitative agreement with these experimental values, and 
particularly, explain that the entropy related to the lower-temperature 
transition is considerably smaller than that for the higher-temperature transition. 
The small entropy related to the transition at $T_{\rm N}$ is likely due to 
the magnetic frustration and a 1D AF correlation well developed 
above $T_{\rm N}$ which will be discussed in Sec.~\ref{Sec:1D}.

\subsubsection{High-temperature transition: Orbital ordering}
\label{Sec:Orbital Ordering}

\begin{figure}
\includegraphics[width=8cm]{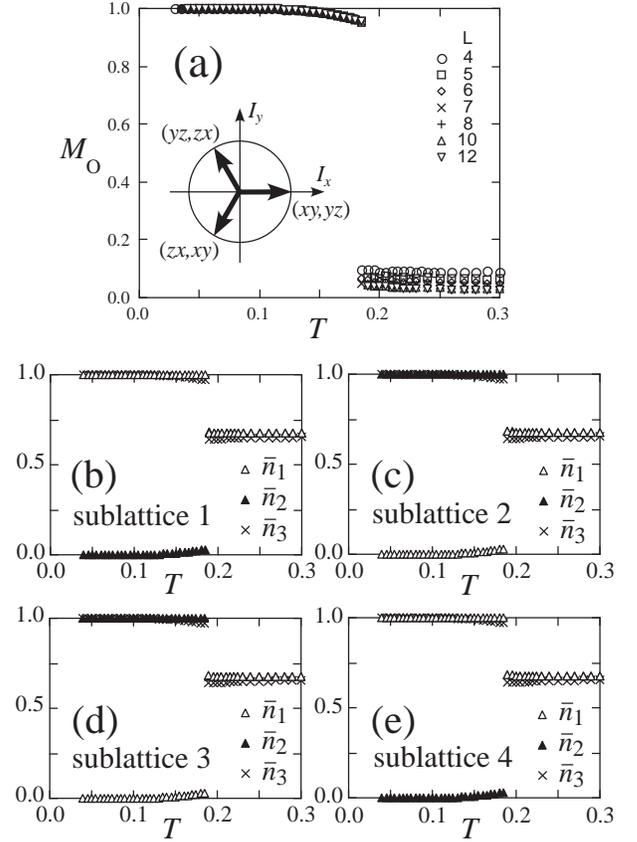}
\caption{
(a) The sublattice orbital moment in Eq.~(\ref{eq:M_O}) at $J_3/J = 0.02$. 
The inset shows the three-state clock vector for the orbital state. 
(b)-(d) Electron density in each orbital for four sublattices $1$-$4$ 
shown in Fig.~\ref{fig:orbital order} (b) for $L=12$ [Eq.~(\ref{eq:nbar})]. 
Error bars are smaller than the symbol sizes. 
} 
\label{fig:orbital} 
\end{figure}

To characterize two transitions in Fig.~\ref{fig:E and C}, 
we calculate corresponding order parameters. 
First, we consider the transition at the higher temperature 
$T_{\rm O} \simeq 0.19J$. 
Figure~\ref{fig:orbital} (a) shows the sublattice orbital moment, 
which is defined in the form 
\begin{equation}
M_{\rm O} = \frac{4}{N_{\rm site}} 
\Big\langle \Big| \sum_{i \in {\rm sublattice}} \mib{I}_i \Big| \Big\rangle, 
\label{eq:M_O}
\end{equation}
where the summation is taken over the sites within one of 
the four sublattices in Fig.~\ref{fig:orbital order} (b). 
Here, $\mib{I}_i$ is the three-state clock vector at the site $i$ which describes 
three different orbital states as shown in the inset of Fig.~\ref{fig:orbital} (a); 
$\mib{I}_i = (1,0)$ for ($xy,yz$), 
$\mib{I}_i = (-1/2,\sqrt3/2)$ for ($yz,zx$), and 
$\mib{I}_i = (-1/2,-\sqrt3/2)$ for ($zx,xy$) orbital occupations, respectively. 
It is found that the values of $M_{\rm O}$ for four different sublattices 
have the same value within the error bars 
so that we omit the sublattice index in Eq.~(\ref{eq:M_O}). 
As shown in Fig.~\ref{fig:orbital} (a), 
$M_{\rm O}$ shows a clear jump 
at the same temperature as for the internal energy and the specific heat. 
This suggests that a four-sublattice orbital ordering occurs at $T_{\rm O}$. 
At low temperatures, $M_{\rm O}$ approaches its maximum value $1$, 
which indicates the four-sublattice orbital order becomes almost perfect there. 

Figures~\ref{fig:orbital} (b)-(e) show the orbital distribution 
for four sublattices $1$-$4$ 
shown in Fig.~\ref{fig:orbital order} (b), respectively, 
which is defined as 
\begin{equation}
\bar{n}_\alpha = \frac{4}{N_{\rm site}} \sum_{i \in {\rm sublattice}} 
\langle n_{i \alpha} \rangle, 
\label{eq:nbar}
\end{equation}
where $\alpha = 1$ ($d_{yz}$), $2$ ($d_{zx}$), and $3$ ($d_{xy}$).
The results indicate that at $T_{\rm O}$ 
the orbital distributions suddenly change 
from equally distributed $\bar{n}_\alpha \sim 2/3$ 
in the para phase above $T_{\rm O}$ 
to almost polarized $\bar{n}_\alpha \sim 0$ or $1$ for $T < T_{\rm O}$. 
In the orbital ordered phase below $T_{\rm O}$, 
$d_{yz}$ ($\alpha=1$) and $d_{xy}$ ($\alpha=3$) orbitals are occupied 
in the sublattices 1 and 4, and 
$d_{zx}$ ($\alpha=2$) and $d_{xy}$ ($\alpha=3$) orbitals are occupied 
in the sublattices 2 and 3 (Ref.~\onlinecite{note3}). 
This orbital ordering structure is shown in Fig.~\ref{fig:orbital structure}. 
This pattern is consistent with the mean-field prediction 
in Fig.~\ref{fig:orbital order} (b). 

\begin{figure}
\caption{
Orbital ordering structure obtained by MC calculations. 
Dark blue and light yellow octahedra in (a) contain V cations 
where ($d_{xy},d_{zx}$) and ($d_{xy},d_{yz}$) orbitals 
are occupied, respectively. 
They stack alternatively in the $z$ direction. 
Some of $d_{zx}$ and $d_{yz}$ orbitals are shown 
by white and black lobes, respectively. 
(b) is the projection of (a) from the $z$ direction. 
$d_{xy}$ orbitals are singly occupied at all the sites and 
not shown in the figures. 
}
\label{fig:orbital structure} 
\end{figure}

\begin{figure}
\includegraphics[width=7cm]{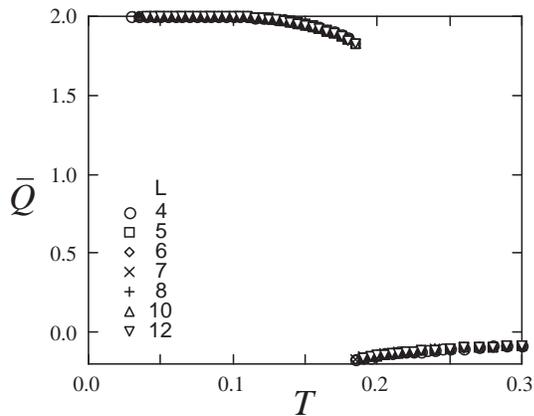}
\caption{
The average of the JT distortion defined in Eq.~(\ref{eq:Qbar}) 
at $J_3/J = 0.02$.
Error bars are smaller than the symbol sizes. 
} 
\label{fig:JT} 
\end{figure}

Accompanying the transition at $T_{\rm O}$, 
a tetragonal JT distortion occurs discontinuously. 
In Fig.~\ref{fig:JT}, we plot 
the average of the JT distortions which is calculated by 
\begin{equation}
\bar{Q} = \sum_i \langle Q_i \rangle / N_{\rm site}. 
\label{eq:Qbar}
\end{equation}
The positive value of $\bar{Q}$ below $T_{\rm O}$ 
corresponds to a ferro-type tetragonal JT distortion 
with a flattening of VO$_6$ octahedra 
as mentioned in Sec.~\ref{Sec:Spin-Orbital Model}. 
Therefore, the level splitting 
shown in Fig.~\ref{fig:orbital order} (a) is realized. 
The value of $\bar{Q}$ approaches 2 at low temperatures 
which is the mean-field value obtained in Appendix A. 
We note that $\bar{Q}$ is small but finite even for $T > T_{\rm O}$. 
This is because $H_{\rm JT}$ breaks the cubic symmetry of the system 
as mentioned before. 

Therefore, the discontinuous phase transition at $T_{\rm O}$ 
is ascribed to the orbital ordering with the pattern of 
Fig.~\ref{fig:orbital structure} 
accompanied by the tetragonal JT distortion 
with the flattening of VO$_6$ octahedra. 
From Figs.~\ref{fig:orbital} and \ref{fig:JT}, 
we estimate $T_{\rm O} = (0.19 \pm 0.01)J$ for $J_3/J=0.02$. 

\subsubsection{Low-temperature transition: Antiferromagnetic Spin Ordering}
\label{Sec:AF Ordering}

\begin{figure}
\includegraphics[width=7cm]{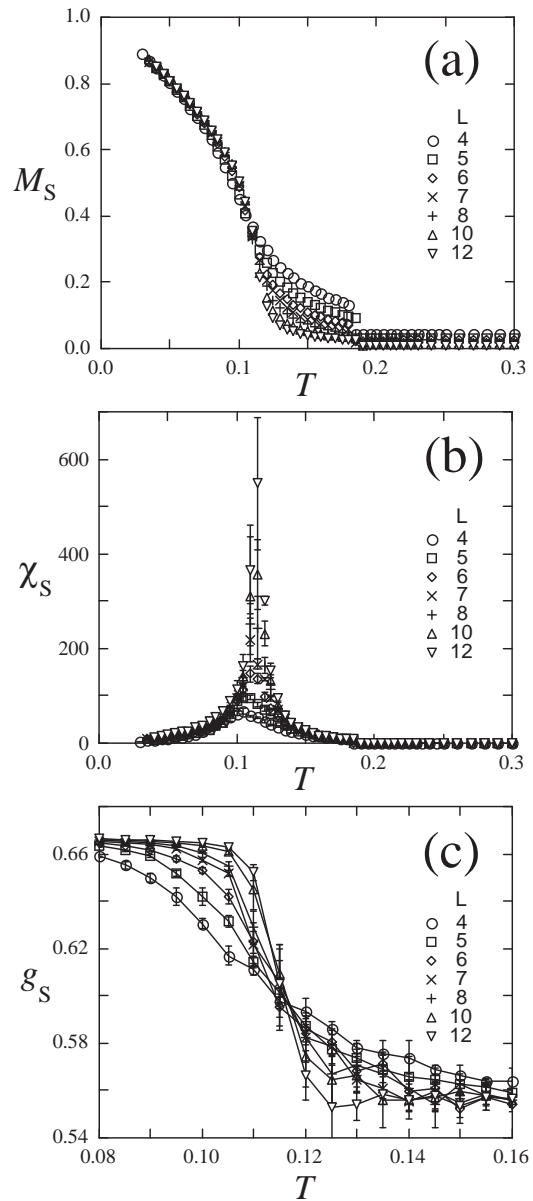}
\caption{
(a) The staggered moment defined in Eq.~(\ref{eq:M_S}). 
Error bars are smaller than the symbol sizes. 
(b) The staggered magnetic susceptibility in Eq.~(\ref{eq:chi_S}). 
(c) The Binder parameter defined in Eq.~(\ref{eq:g_S}). 
The lines are guides for the eyes. 
All the results are calculated at $J_3/J = 0.02$.
} 
\label{fig:spin} 
\end{figure}

Next, we consider the other transition at $T_{\rm N} \simeq 0.115J$. 
Figure~\ref{fig:spin} (a) shows the temperature dependence of 
the staggered magnetization defined in the form 
\begin{equation}
M_{\rm S} = \langle |\mib{f}|^2 \rangle^{1/2}, 
\label{eq:M_S}
\end{equation}
where the structure factor is given by 
\begin{equation}
\mib{f} = \frac{2}{N_{\rm site}} \sum_{i_{\rm ch}=1}^{N_{\rm ch}} 
\exp(2 \pi i l_{i_{\rm ch}}^z) \big[ {\sum_{i}}' \mib{S}_i (-1)^{4 y_i} \big]. 
\label{eq:f}
\end{equation}
This definition looks complicated but nothing but 
the order parameter of the spin ordering pattern 
shown in Fig.~\ref{fig:spin order}. 
Here, the first summation is taken over 
different chains lying in the $xy$ planes 
($N_{\rm ch}$ is the total number of the $xy$ chains in the system, 
i.e., $N_{\rm ch} = 4 L^2$ 
where $L$ is the linear dimension of the system 
measured in the cubic units), and 
the second summation $\sum'$ is taken over the sites 
in the $i_{\rm ch}$-th $xy$ chain. 
$y_i$ is the $y$ coordinates of the site $i$, and 
$l_{i_{\rm ch}}^z$ is the $z$ coordinate of the $i_{\rm ch}$-th $xy$ chain 
measured in the cubic units. 
We set the normalization in Eq.~(\ref{eq:f}) such that $M_{\rm S}$ becomes $1$ 
for the fully saturated AF order in Fig.~\ref{fig:spin order}. 
Note that the structure factor $\mib{f}$ cannot be defined 
only by a real phase factor 
because of the complicated AF ordering pattern which is expected 
to have the four-period structure in the $yz$ and $zx$ directions 
as shown in Fig.~\ref{fig:spin order}. 
See also the discussions in Sec.~\ref{Sec:Collinearity}. 
The structure factor is also expressed in a simpler form as
\begin{equation}
\mib{f} = \frac{2}{N_{\rm site}} \sum_i g_i \mib{S}_i, 
\end{equation}
where the form factor $g_i$ is given by
\begin{equation}
g_i = \cos[2\pi (x_i + y_i)] + i \cos[2\pi (x_i - y_i)]. 
\end{equation}
Note that $g_i$ is specified only by the $x$ and $y$ coordinates of the site $i$ 
because the $z$ coordinate is uniquely determined within the cubic unit cell 
due to the special structure of the pyrochlore lattice. 
[See the projection in Fig.~\ref{fig:spin order} (b).]

As shown in Fig.~\ref{fig:spin} (a), 
the staggered magnetization $M_{\rm S}$ develops continuously 
below $T_{\rm N} \simeq 0.115J$, and 
approaches the fully saturated value $M_{\rm S} = 1$ 
as $T \rightarrow 0$. 
Figure~\ref{fig:spin} (b) shows the staggered magnetic susceptibility obtained by 
\begin{equation}
\chi_{\rm S} = \frac{N_{\rm site}}{T}
\big(\langle |\mib{f}|^2 \rangle 
- |\langle \mib{f} \rangle|^2 \big). 
\label{eq:chi_S}
\end{equation}
Here, we calculate the susceptibility 
by using $\langle |\mib{f}| \rangle$ 
instead of $| \langle \mib{f} \rangle|$ in Eq.~(\ref{eq:chi_S}) 
for convenience of numerical calculations 
since both quantities agree with each other in the thermodynamic limit. 
The susceptibility $\chi_{\rm S}$ shows a diverging behavior at $T_{\rm N}$ and 
the peak value increases with the system size. 
In Fig.~\ref{fig:spin} (c), we show the Binder parameter 
for the staggered magnetization which is defined as 
\cite{Binder1981} 
\begin{equation}
g_{\rm S} = 1 - \frac{\langle (|\mib{f}|^2)^2 \rangle} 
{3 \langle |\mib{f}|^2 \rangle^2}. 
\label{eq:g_S} 
\end{equation}
It is known that the Binder parameter becomes larger (smaller) 
for larger system sizes in the ordered (disordered) phase, and hence, 
the crossing point of the Binder parameter for different system sizes 
gives a good estimate of the transition temperature. 
The MC data in Fig.~\ref{fig:spin} (c) shows the crossing 
at the same temperature of the divergence of $\chi_{\rm S}$ 
in Fig.~\ref{fig:spin} (b), 
which indicates the phase transition 
by the order parameter $M_{\rm S}$ at $T_{\rm N}$. 

All these results in Fig.~\ref{fig:spin} 
indicate that the phase transition at $T_{\rm N}$ 
is caused by the continuous growth of the staggered magnetic order. 
From Figs.~\ref{fig:spin} (b) and (c), 
we estimate $T_{\rm N} = (0.115 \pm 0.005)J$ for $J_3/J=0.02$. 
The magnetic ordering pattern will be discussed in the next section. 

\subsubsection{Collinearity of magnetic ordering}
\label{Sec:Collinearity}

The mean-field argument in Ref.~\onlinecite{Tsunetsugu2003} 
and in Sec.~\ref{Sec:Spin-Orbital Model} predicts that 
the staggered magnetizations develop independently on two sublattices 
and 
the relative angle between two AF moments is free. 
The two sublattices are shown in Fig.~\ref{fig:two sublattices}; 
one consists of the $[110]$ chains and the other consists of the $[1\bar1 0]$ chains.  
The staggered magnetization $M_{\rm S}$ defined in Eq.~(\ref{eq:M_S}) 
is the order parameter of the 3D AF spin ordering in Fig.~\ref{fig:spin order}, 
but cannot measure the collinearity between two staggered magnetizations. 
For example, $M_{\rm S}$ takes the value of $1$ 
independent of the relative angle between staggered moments 
in two sublattices, $\mib{M}_1$ and $\mib{M}_2$, 
shown in Fig.~\ref{fig:two sublattices}, 
when both sublattice moments saturate. 
The spin configuration in Fig.~\ref{fig:two sublattices} is 
a noncollinear one, while that in Fig.~\ref{fig:spin order} is 
a collinear one, and both give $M_{\rm S} = 1$.  

Here, we measure the collinearity between 
the staggered moments in the two sublattices by 
\begin{equation}
C_{12} = \Big\langle \frac{(\mib{M}_1 \cdot \mib{M}_2)^2}
{(\mib{M}_1)^2 (\mib{M}_2)^2} \Big\rangle 
= \langle \cos^2 \theta_{12} \rangle. 
\label{eq:C12}
\end{equation}
Here we may consider $\theta_{12}$ the angle 
between $\mib{M}_1$ and $\mib{M}_2$. 
In the present calculations, 
the two sublattice moments are obtained 
by the real and imaginary parts of the structure factor $\mib{f}$ 
in Eq.~(\ref{eq:f}) as
\begin{equation}
\mib{M}_1 = \mbox{Re}(\mib{f}), \ 
\mib{M}_2 = \mbox{Im}(\mib{f}), 
\label{eq:M1&M2}
\end{equation}
respectively. 
If the AF order is collinear, i.e., $\mib{M}_1 \parallel \mib{M}_2$, 
the value of $C_{12}$ become $1$. 
Figure~\ref{fig:C12} shows the MC results. 
Below $T_{\rm N} \simeq 0.115J$, $C_{12}$ is rapidly enhanced, and 
becomes larger and approaches $1$ as the system size increases.  
This indicates that the magnetic state is collinear below $T_{\rm N}$. 
Thus, the AF order below $T_{\rm N}$ is identified 
by the collinear AF spin order 
whose pattern is given by Fig.~\ref{fig:spin order}. 

\begin{figure}
\caption{
Two sublattices; one consists of the [$110$] chains (blue dashed lines) and 
the other consists of the [$1\bar1 0$] chains (red solid lines). 
In each sublattice, chains are connected 
by the third-neighbor exchange $J_3$ 
in the $yz$ and $zx$ directions. 
Black (white) arrows 
show spins in the sublattice moment $\mib{M}_1$ ($\mib{M}_2$). 
} 
\label{fig:two sublattices} 
\end{figure}

\begin{figure}
\includegraphics[width=7cm]{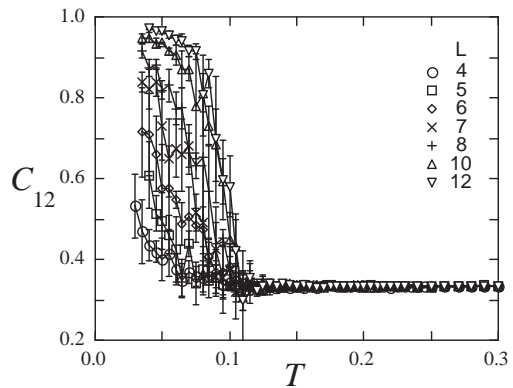}
\caption{
The collinearity defined in Eq.~(\ref{eq:C12}) at $J_3/J = 0.02$. 
The lines are guides for the eyes. 
} 
\label{fig:C12} 
\end{figure}

The result indicates that thermal fluctuations 
in the classical version of the model (\ref{eq:H}) lift 
the degeneracy of the relative angle between two sublattice magnetizations 
and stabilize the collinear spin structure. 
This is a kind of the so-called order-by-disorder phenomena. 
\cite{Villain1977} 
In Sec.~\ref{Sec:Spin-Orbital Model} and Ref.~\onlinecite{Tsunetsugu2003}, 
we discussed that the collinear order may appear 
also by quantum fluctuations. 
Thus, we can conclude that both thermal and quantum fluctuations 
favor a collinear state of the two sublattice magnetizations. 
It is known that 
in many frustrated systems, thermal and quantum fluctuations favor a collinear state. 
This is also the case for the model (\ref{eq:H}). 

\subsubsection{Uniform magnetic susceptibility}
\label{Sec:chi}

\begin{figure}
\includegraphics[width=7cm]{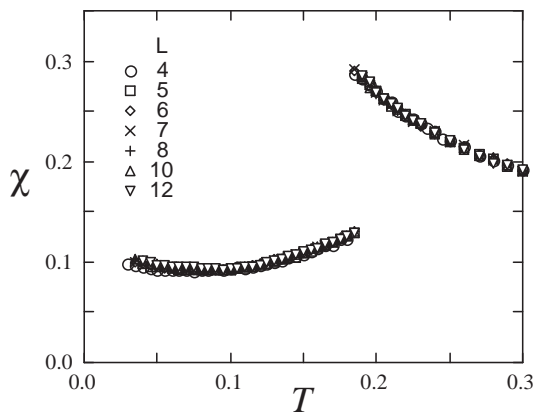}
\caption{
The uniform magnetic susceptibility in Eq.~(\ref{eq:chi}) at $J_3/J = 0.02$. 
Error bars are smaller than the symbol sizes. 
} 
\label{fig:chi} 
\end{figure}

We show in Fig.~\ref{fig:chi} the temperature dependence of 
the uniform magnetic susceptibility calculated by 
\begin{equation}
\chi = \frac{N_{\rm site}}{T}
(\langle M_{\rm tot}^2 \rangle - \langle M_{\rm tot} \rangle^2), 
\label{eq:chi}
\end{equation}
where $M_{\rm tot}$ is the total magnetic moment per site, 
$
M_{\rm tot} = | \sum_i \mib{S}_i | / N_{\rm site}. 
$
The result corresponds to the zero-field-cool (ZFC) result in experiments 
rather than the field-cool (FC) result 
since the susceptibility is measured by starting the MC simulation 
from an initial state with zero magnetic field at each temperature
as described in Sec.~\ref{Sec:Monte Carlo}. 
The MC results show a sudden drop at $T_{\rm O} \simeq 0.19J$ and 
a little continuous change at $T_{\rm N} \simeq 0.115J$. 
These features qualitatively agree with the experimental results. 
\cite{Ueda1997} 

In experiments, a large difference between ZFC and FC results 
has been found. 
The difference develops from well above $T_{\rm c1}$, and 
remains substantial even below $T_{\rm c2}$. 
We will comment on this behavior in Sec.~\ref{Sec:ZFC and FC}.

\subsection{One-dimensional spin correlation in the intermediate phase}
\label{Sec:1D}

As shown in Fig.~\ref{fig:spin} (a), 
$M_{\rm S}$ in finite-size systems are enhanced below $T_{\rm O}$ 
even above $T_{\rm N}$. 
In this section, we show that there the short-range AF correlation 
is much enhanced along 1D chains in the $xy$ planes 
compared to the $yz$ and $zx$ chains.  

\begin{figure}
\includegraphics[width=7cm]{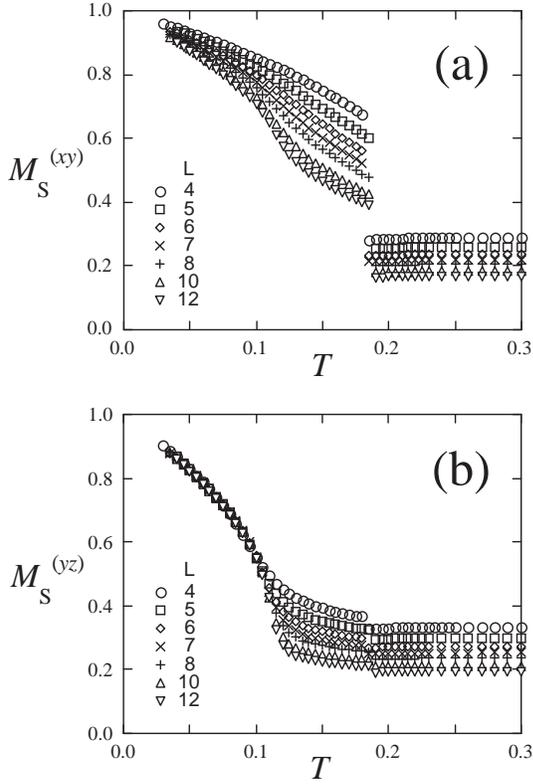}
\caption{
The staggered magnetic moment along 
(a) the $xy$ chains [Eq.~(\ref{eq:M_S^xy})] and 
(b) the $yz$ chains [Eq.~(\ref{eq:M_S^yz})] at $J_3/J = 0.02$. 
Error bars are smaller than the symbol sizes. 
} 
\label{fig:M_S chains vs T} 
\end{figure}

The mean-field arguments in Sec.~\ref{Sec:Spin-Orbital Model} and 
in Ref.~\onlinecite{Tsunetsugu2003} suggest that 
the orbital order enhances 1D AF correlations within the $xy$ chains 
in the intermediate phase $T_{\rm N} < T < T_{\rm O}$. 
To examine the spatially anisotropic spin correlation, 
we calculate the staggered magnetic moments along the chains 
in different directions. 
The staggered moment along the $xy$ chains may be defined by 
\begin{equation}
M_{\rm S}^{(xy)} = \Big\langle \Big[ \frac{1}{N_{\rm ch}} 
\sum_{i_{\rm ch} = 1}^{N_{\rm ch}} 
\big| \frac{1}{4L} 
{\sum_i}' \mib{S}_i \exp(4 \pi i x_i)
\big|^2 \Big]^{1/2} \Big\rangle, 
\label{eq:M_S^xy}
\end{equation}
where the summations are taken in the same manner as in Eq.~(\ref{eq:f}), and 
the phase factor describes the $\uparrow$-$\downarrow$-$\uparrow$-$\downarrow$-
$\cdot\cdot\cdot$ structure in the $xy$ chains 
as shown in Fig.~\ref{fig:spin order}. 
Similarly, the ``staggered" moment along the $yz$ chains is calculated by 
\begin{equation}
M_{\rm S}^{(yz)} = \Big\langle \Big[ \frac{2}{N_{\rm ch}} 
\sum_{i_{\rm ch} = 1}^{N_{\rm ch}} 
\big| \frac{1}{4L} 
{\sum_i}' \mib{S}_i \exp(2 \pi i z_i) 
\big|^2 \Big]^{1/2} \Big\rangle, 
\label{eq:M_S^yz}
\end{equation}
where the first summation is taken over all the $yz$ chains in the system and 
the second summation is taken over the sites within the $i_{\rm ch}$-th $yz$ chain 
with a phase factor of period four in the $z$ direction 
with considering the $\uparrow$-$\uparrow$-$\downarrow$-$\downarrow$-
$\cdot\cdot\cdot$ structure as shown in Fig.~\ref{fig:spin order}. 
Here, $N_{\rm ch}$ is the number of the $xy$ or $yz$ chains, 
i.e., $N_{\rm ch} = 4L^2$ where $L$ is the linear dimension 
of the system measured in the unit cell. 
The normalization factors in Eqs.~(\ref{eq:M_S^xy}) and (\ref{eq:M_S^yz}) 
are given such that both $M_{\rm S}^{(xy)}$ and $M_{\rm S}^{(yz)}$ 
become $1$ in the fully saturated AF state with 
$|\mib{M}_1| = |\mib{M}_2| = 1$ in Eq.~(\ref{eq:M1&M2}). 

Figure~\ref{fig:M_S chains vs T} shows the MC results. 
The moment in the $zx$ chains, $M_{\rm S}^{(zx)}$, has the same value 
as $M_{\rm S}^{(yz)}$ within the statistical error bars. 
Below $T_{\rm O} \simeq 0.19J$, 
the moment in the $xy$ chains, $M_{\rm S}^{(xy)}$, 
is much more enhanced compared to that in the $yz$ and $zx$ chains, 
$M_{\rm S}^{(yz)}$ and $M_{\rm S}^{(zx)}$. 
This indicates that in the intermediate phase $T_{\rm N} < T < T_{\rm O}$, 
the AF spin correlations develop mainly within the $xy$ chains. 

\begin{figure}
\includegraphics[width=8cm]{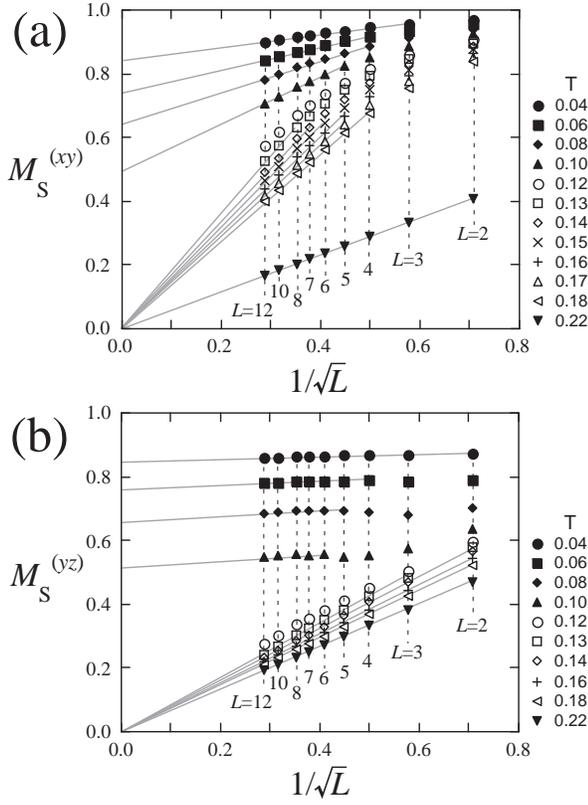}
\caption{
System-size extrapolations of the staggered magnetic moment along 
(a) the $xy$ chains and (b) the $yz$ chains at $J_3/J = 0.02$. 
Error bars are smaller than the symbol sizes. 
The gray lines are the fits to a linear function of $1/\sqrt{L}$. 
See the text for details. 
} 
\label{fig:M_S chain vs L^-0.5} 
\end{figure}

The system-size dependence of these moments within the chains 
provides further information. 
Figure~\ref{fig:M_S chain vs L^-0.5} plots the system-size extrapolations 
of the MC data in Fig.~\ref{fig:M_S chains vs T}. 
In the ordered phase below $T_{\rm N} \simeq 0.115J$, 
the MC data are extrapolated to finite values, which 
indicates a long-range order. 
The extrapolated values of $M_{\rm S}^{(xy)}$ and $M_{\rm S}^{(yz)}$ 
to $L \to \infty$ 
agree with each other within the error bars as expected in this 3D ordered phase. 
On the other hand, in the disordered phase above $T_{\rm O} \simeq 0.19J$, 
the MC data for both $M_{\rm S}^{(xy)}$ and $M_{\rm S}^{(yz)}$ 
well scale to $1/\sqrt{L}$ and are extrapolated to zero. 
(Only the data at $T=0.22J$ are shown in the figures as a typical example 
but other data at $T > T_{\rm O}$ show a similar behavior.)
This scaling implies the exponential decay of the spin-spin correlation 
along the chain as explained below. 
Compared to the rather isotropic behavior for 
$T < T_{\rm N}$ and $T > T_{\rm O}$, 
we find a contrasting behavior 
between $M_{\rm S}^{(xy)}$ and $M_{\rm S}^{(yz)}$ 
in the intermediate phase $T_{\rm N} < T < T_{\rm O}$; 
almost all the data of $M_{\rm S}^{(yz)}$ well scale to $1/\sqrt{L}$ 
even for the smallest system size in the present calculations with $L=2$, 
while the data of $M_{\rm S}^{(xy)}$ show a clear deviation from the scaling 
in the small-$L$ region. 

In order to analyze this contrasting behavior 
in the intermediate phase, 
we introduce the correlation length $\xi$ along the chain, 
and assume a simple scaling form of the spin-spin correlation function; 
the amplitude of staggered spin-spin correlation becomes an almost constant value 
within the length scale $\xi$ and decays exponentially 
for further distance than $\xi$, 
that is, 
\begin{eqnarray}
| \langle \mib{S}_i \cdot \mib{S}_j \rangle | 
&\sim& c \quad {\rm for} \ \ r_{ij} < \xi
\nonumber \\
&\sim& c \exp[ -(r_{ij}-\xi)/\xi] \quad {\rm for} \ \ r_{ij} > \xi,
\label{eq:SSscaling}
\end{eqnarray}
where $c$ is a constant and $r_{ij} = | \mib{r}_i - \mib{r}_j |$ is the distance 
between sites $i$ and $j$ along the chain 
measured in the cubic units. 
The staggered moments in Eqs.~(\ref{eq:M_S^xy}) and (\ref{eq:M_S^yz}) can be 
rewritten by using the spin-spin correlation. 
For instance, Eq.~(\ref{eq:M_S^xy}) is rewritten in the form 
\begin{eqnarray}
M_{\rm S}^{(xy)} &\sim& \Bigg[ 
\frac{1}{(4L)^2} {\sum_{ij}}' \langle \mib{S}_i \cdot 
\mib{S}_j 
\cos 4\pi (x_i + y_i)
\rangle \Bigg]^{1/2}
\nonumber
\\
&\sim& \Bigg[ \frac{1}{\sqrt2 L} \int \big| \langle \mib{S}_0 \cdot 
\mib{S}_r \rangle \big| dr \Bigg]^{1/2}, 
\label{eq:M_S^xy SS}
\end{eqnarray} 
by introducing the continuum limit to evaluate the summation. 
Note that $\sqrt2 L$ is the length of the chains in the $L^3$ system. 
Therefore, by assuming the scaling form in Eq.~(\ref{eq:SSscaling}), 
we obtain 
\begin{equation}
M_{\rm S}^{(xy)} \sim 
\Bigg[ \frac{1}{\sqrt2 L} \int_0^{\sqrt2 L} c \ dr \Bigg]^{1/2}
= \sqrt{c}
\label{eq:scaling near}
\end{equation}
for the case of $\xi > \sqrt2 L$, and 
\begin{eqnarray}
M_{\rm S}^{(xy)} &\sim& 
\Bigg[ \frac{1}{\sqrt2 L} \Big\{ \int_0^\xi c \ dr 
+ ce \int_\xi^{\sqrt2 L} \exp\big( -\frac{r}{\xi} \big)  dr 
\Big\} \Bigg]^{1/2} 
\nonumber \\ 
&=& \Bigg[ \frac{1}{\sqrt2 L} \Big\{ 2c\xi 
- ce\xi \exp\big(-\frac{\sqrt2 L}{\xi} \big) \Big\} \Bigg]^{1/2}
\nonumber \\ 
&\sim& \frac{2^{1/4} \sqrt{c\xi}}{\sqrt{L}} 
\label{eq:M_S^xy SS2}
\label{eq:scaling far}
\end{eqnarray}
for the case of $\xi < \sqrt2 L$. 
The staggered moment in the $yz$ chains in Eq.~(\ref{eq:M_S^yz}) is 
also estimated in a similar manner. 

With considering the above arguments based on the scaling form (\ref{eq:SSscaling}), 
we can roughly estimate $\xi$ 
from the crossover of the scaling 
from $1/\sqrt{L}$ [Eq.~(\ref{eq:scaling far})] 
to the constant [Eq.~(\ref{eq:scaling near})] (+ corrections) 
in the system-size dependence in Fig.~\ref{fig:M_S chain vs L^-0.5}. 
That is, the linear dimension of the system size where the crossover occurs 
gives a rough estimate of $\xi$. 
For instance, at $T = 0.13J$, $M_{\rm S}^{(xy)}$ deviates from 
the scaling of $1/\sqrt{L}$ at $L \simle 8$, on the contrary, 
$M_{\rm S}^{(yz)}$ obeys the scaling down to $L \sim 3$. 
This suggests that $\xi^{(xy)} \sim 6$ ($\sim 16$ sites) 
and $\xi^{(yz)} \sim 2$ ($\sim 6$ sites). 
[Note that the longest distance along the chain in the periodic $L^3$ system 
($N_{\rm site} = 16 L^3$) is $L/\sqrt{2}$.] 
The correlation length in the $xy$ direction is about 3 times longer 
than in the $yz$ and $zx$ directions. 
We note that the crossover length systematically 
shifts to a longer value as decreasing temperature, 
which suggests divergence of $\xi$ as $T \to T_{\rm N}$. 
[$\xi$ is expected to diverge in all the directions 
with the same critical exponent 
although the divergence of $\xi^{(yz)}$ is not clear 
in Fig.~\ref{fig:M_S chain vs L^-0.5} (b)
compared to $\xi^{(xy)}$.]

Therefore, we find a 1D anisotropy of the spin correlation: 
The magnitude of the moment is much enhanced and 
the correlation length is much longer in the $xy$ chains 
than in the $yz$ and $zx$ chains. 
As predicted in the mean-field arguments, 
the staggered spin correlation develops dominantly in the $xy$ chains due to 
the cooperation of the orbital ordering and the geometrical frustration. 

Recently, the neutron scattering experiment has been performed 
in both above and below $T_{\rm c1}$. 
\cite{LeePREPRINT} 
A clear difference of the $Q$ dependence of the inelastic neutron intensity 
has been found between the data above and below $T_{\rm c1}$, 
and ascribed to the 1D spin anisotropy due to the orbital ordering.  
The proposed pattern of the orbital order is 
the same as our result in Sec.~\ref{Sec:Orbital Ordering}.

\subsection{Anisotropy in optical response}
\label{Sec:Optical Response}

It has been recognized that the interplay between orbital and spin 
degrees of freedom often causes an anisotropic electronic state, 
which is observed in optical measurements. 
For instance, in perovskite vanadium oxides, 
a strong 1D nature in the 3D lattice structure 
has been observed in the optical measurement, 
\cite{Miyasaka2002} 
and theoretically ascribed to the anisotropic 
orbital exchange in the spin ordered state. 
\cite{Motome2003} 
Therefore, we expect that 
the 1D anisotropy found in the previous section 
also shows up in the optical response. 

We consider the anisotropy of the optical response 
by calculating the spectral weight in the present spinel case. 
The spectral weight is the total weight of the optical conductivity 
defined by 
\begin{equation}
I_\mu = \frac{2 \hbar^2}{\pi e^2} \int_0^\infty \sigma_{\mu \mu} (\omega) d\omega, 
\label{eq:I_mu}
\end{equation}
where $e$ is the electron charge, $\hbar = h/2\pi$ is the Planck's constant, 
$\sigma_{\mu \mu}$ is the diagonal element of the optical conductivity tensor, 
and $\mu = x, y,$ or $z$. 
Here, we consider only the optical transfer 
within the $3d$ $t_{2g}$ electron bands and 
neglect that to the other bands such as the $3d$ $e_g$ bands and 
the oxygen $p$ bands. 
In experiments, the integral of Eq.~(\ref{eq:I_mu}) is calculated 
up to an appropriate energy cutoff to extract the spectral weight from 
the relevant $t_{2g}$ levels. 
The spectral weight in Eq.~(\ref{eq:I_mu}) is generally given 
by the kinetic energy in the $t_{2g}$ Hubbard model (\ref{eq:H_multiorbital}). 
\cite{Maldague1977,Shastry1990} 
In the strong correlation limit, it is calculated 
by the spin and orbital exchange energy 
in the effective model (\ref{eq:H_SO}). 
\cite{Motome2003} 
We show in the following only the results of the calculations. 
Details of the calculations are explained in Appendix B. 

In the calculation of the spectral weight for the present model, 
there are three points to be noticed. 
One is that the original Hubbard model (\ref{eq:H_multiorbital}) 
contains not only the nearest-neighbor 
but also the third-neighbor hoppings. 
These two types of hopping contribute differently to the spectral weight. 
The second point is that the directions of the electron hoppings 
are different from the crystal axes. 
Electrons hop along the $xy$, $yz$, and $zx$ chains 
which are canted by 45 degree from two crystal axes and 
perpendicular to the rest one. 
From this, for instance, $I_z$ is given by the kinetic energy 
both in the $zx$ and $yz$ chains but not in the $xy$ chains. 
The third point is the crystal symmetry. 
Since the system shows the tetragonal lattice distortion below $T_{\rm O}$, 
there holds a general relation in the form 
\begin{equation}
I_x = I_y \neq I_z
\label{eq:Ix,Iy,Iz}
\end{equation}
for $T < T_{\rm O}$. 
We note that there remains weak anisotropy of Eq.~(\ref{eq:Ix,Iy,Iz})
even in the high-temperature phase 
above $T_{\rm O}$ because of the broken cubic symmetry in $H_{\rm JT}$. 
With taking account of these points, 
we can derive the following expressions 
\begin{eqnarray}
I_x &=& \frac{-1}{2N_{\rm site}} \sum_{\zeta=zx,xy} \Big[ 
\big\langle (H_{\rm SO}^{\rm nn})_{\zeta} \big\rangle 
+ 4 \big\langle (H_{\rm SO}^{\rm 3rd})_{\zeta} \big\rangle 
\Big], 
\label{eq:I_x} 
\\
I_y &=& \frac{-1}{2N_{\rm site}} \sum_{\zeta=xy,yz} \Big[ 
\big\langle (H_{\rm SO}^{\rm nn})_{\zeta} \big\rangle 
+ 4 \big\langle (H_{\rm SO}^{\rm 3rd})_{\zeta} \big\rangle 
\Big], 
\label{eq:I_y} 
\\
I_z &=& \frac{-1}{2N_{\rm site}} \sum_{\zeta=yz,zx} \Big[ 
\big\langle (H_{\rm SO}^{\rm nn})_{\zeta} \big\rangle 
+ 4 \big\langle (H_{\rm SO}^{\rm 3rd})_{\zeta} \big\rangle 
\Big], 
\label{eq:I_z} 
\end{eqnarray} 
where we set $e = \hbar = 1$, and $(H_{\rm SO}^{\rm nn})_{\zeta}$ and 
$(H_{\rm SO}^{\rm 3rd})_{\zeta}$ represent the matrix elements of 
Eqs.~(\ref{eq:H_nn}) and (\ref{eq:H_3rd}) 
in the $\zeta$ chains, respectively ($\zeta = yz$, $zx$, or $xy$). 
The derivation of Eqs.~(\ref{eq:I_x})-(\ref{eq:I_z}) will be given 
in Appendix B. 

In Fig.~\ref{fig:spectral weight}, we show the MC results 
for the spectral weight. 
The spectral weight becomes highly anisotropic 
in the tetragonal phase below $T_{\rm O}$ 
with satisfying the relation (\ref{eq:Ix,Iy,Iz}): 
The spectral weight in the $z$ direction, $I_z$, is 
suddenly suppressed at $T_{\rm O}$, 
while those in the $xy$ plane, $I_x$ and $I_y$, 
are slightly enhanced there. 
Moreover, in the orbital ordered phase below $T_{\rm O}$, 
$I_x$ and $I_y$ increase monotonically with decreasing temperature, 
whereas $I_z$ shows a complicated temperature dependence 
although the dependence itself is small; 
$I_z$ slightly decreases above $T_{\rm N}$, and 
it turns to increase below $T_{\rm N}$. 
The increase below $T_{\rm N}$ comes from 
the energy gain in the $yz$ and $zx$ directions 
by the development of the AF spin order. 

\begin{figure}
\includegraphics[width=7cm]{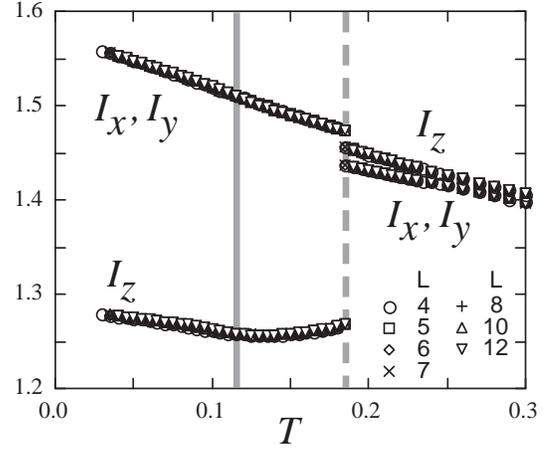}
\caption{
The spectral weights for the $x$, $y$, and $z$ direction 
calculated by Eqs.~(\ref{eq:I_x})-(\ref{eq:I_z}) at $J_3/J = 0.02$. 
Error bars are smaller than the symbol sizes. 
The solid (dashed) line shows $T_{\rm N}$ ($T_{\rm O}$). 
} 
\label{fig:spectral weight} 
\end{figure}

The anisotropic electronic state indicated by our MC results 
can be observed in the optical measurements 
with applying the electric field in each direction. 
It needs a clean surface in the single crystal. 
Unfortunately, it is difficult to grow 
a single crystal large enough thus far, and  
the experimental confirmation of our results 
remains for further study.

\subsection{Systematic changes for $J_3 / J$ and phase diagram}
\label{Sec:Phase Diagram}

\begin{figure}
\includegraphics[width=8cm]{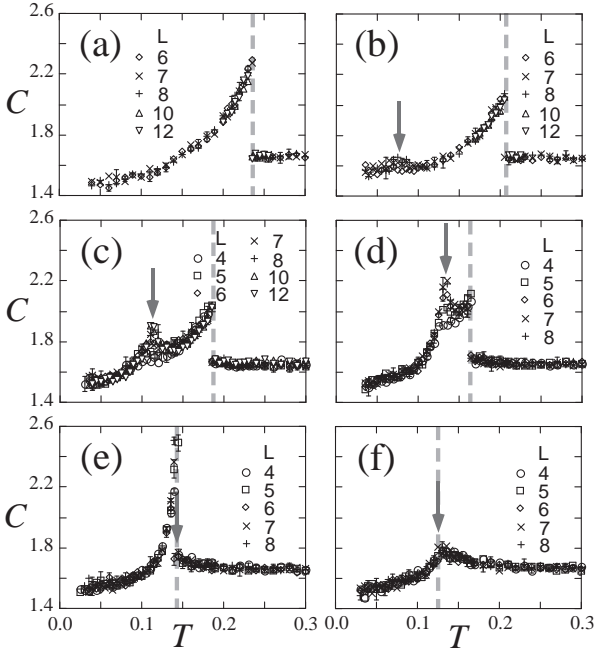}
\caption{
Temperature dependences of the specific heat
at (a) $J_3/J = 0.0$, (b) $0.01$, (c) $0.02$, (d) $0.03$, 
(e) $0.04$, and (f) $0.05$, respectively. 
The dashed lines (the downward arrows) indicate $T_{\rm O}$ ($T_{\rm N}$). 
Typical error bars are shown. 
} 
\label{fig:J_3 dep of C} 
\end{figure}

\begin{figure}
\includegraphics[width=8cm]{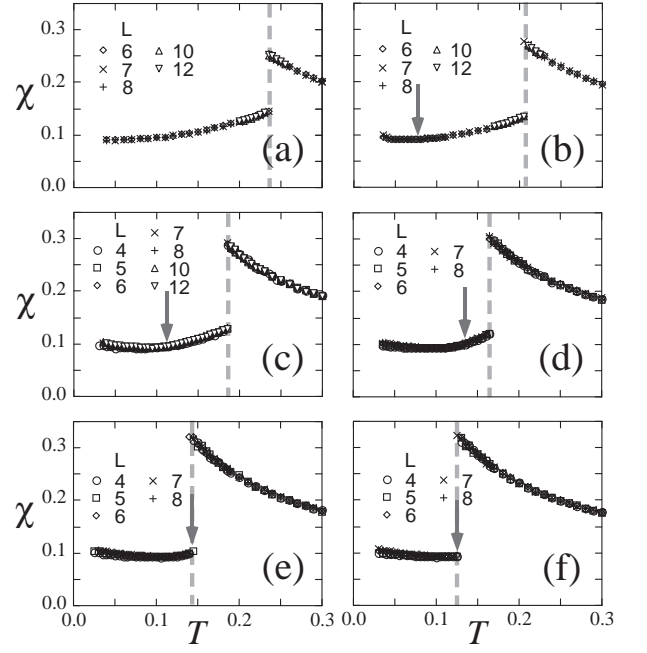}
\caption{
Temperature dependences of the uniform magnetic susceptibility 
at (a) $J_3/J = 0.0$, (b) $0.01$, (c) $0.02$, (d) $0.03$, 
(e) $0.04$, and (f) $0.05$, respectively.
The dashed lines (the downward arrows) indicate $T_{\rm O}$ ($T_{\rm N}$). 
Error bars are smaller than the symbol sizes. 
} 
\label{fig:J_3 dep of chi} 
\end{figure}

\begin{figure}
\includegraphics[width=8cm]{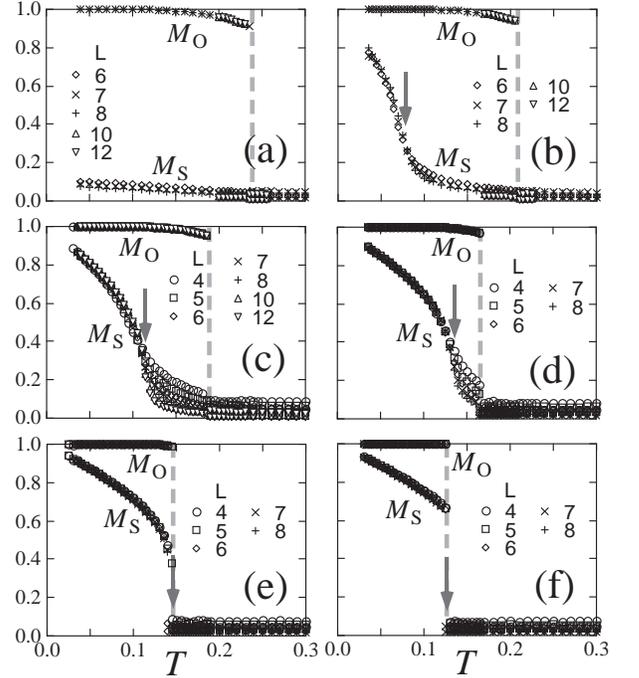}
\caption{
Temperature dependences of the orbital sublattice moment and 
the staggered magnetic moment 
at (a) $J_3/J = 0.0$, (b) $0.01$, (c) $0.02$, (d) $0.03$, 
(e) $0.04$, and (f) $0.05$, respectively.
The dashed lines (the downward arrows) indicate $T_{\rm O}$ ($T_{\rm N}$). 
Error bars are smaller than the symbol sizes. 
} 
\label{fig:J_3 dep of M_O and M_S} 
\end{figure}

Here, we discuss systematic changes with respect to the value of $J_3/J$. 
Figures~\ref{fig:J_3 dep of C}, \ref{fig:J_3 dep of chi}, and 
\ref{fig:J_3 dep of M_O and M_S} show 
the specific heat in Eq.~(\ref{eq:C}), 
the uniform magnetic susceptibility in Eq.~(\ref{eq:chi}), 
and the orbital and magnetic moments 
in Eqs.~(\ref{eq:M_O}) and (\ref{eq:M_S}), respectively, 
for several values of $J_3/J$.  
The orbital and JT transition temperatures $T_{\rm O}$ 
are indicated by the dashed lines in the figures, which 
are monitored by discontinuous changes of $C$, $\chi$, and $M_{\rm O}$ 
as well as $E$ and $\bar{Q}$ (not shown here). 
The AF transition temperatures $T_{\rm N}$ 
are indicated by the downward arrows in the figures, which 
are determined by the singular peak of $C$ and 
a continuous development of $M_{\rm S}$ 
as well as the diverging peak of $\chi_{\rm S}$ and 
the crossing point of $g_{\rm S}$ (not shown here). 

In the case of $J_3 = 0$, the system shows only the orbital and lattice transition, 
and there is no AF ordering down to the lowest temperature. 
Upon switching on $J_3$, the magnetic transition appears at a finite temperature
$T_{\rm N}$, below which the 3D AF order exists. 
As shown in Figs.~\ref{fig:J_3 dep of C}-\ref{fig:J_3 dep of M_O and M_S}, 
$T_{\rm N}$ increases whereas $T_{\rm O}$ slightly decreases as $J_3/J$ increases. 
The increase of $T_{\rm N}$ is easily understood because the AF spin ordering 
is stabilized by the third-neighbor exchange interaction $J_3$. 
The reason for the decrease of $T_{\rm O}$ is not so clear, but 
it might be due to the strong interplay between spin and orbital degrees of freedom 
as well as the frustration between $J$ and $J_3$ within the same chains. 
For $J_3/J \simge 0.04$, $T_{\rm N}$ coincides with $T_{\rm O}$, and 
there is only one discontinuous transition where 
both orbital and AF spin moments become finite discontinuously. 

\begin{figure}
\includegraphics[width=7cm]{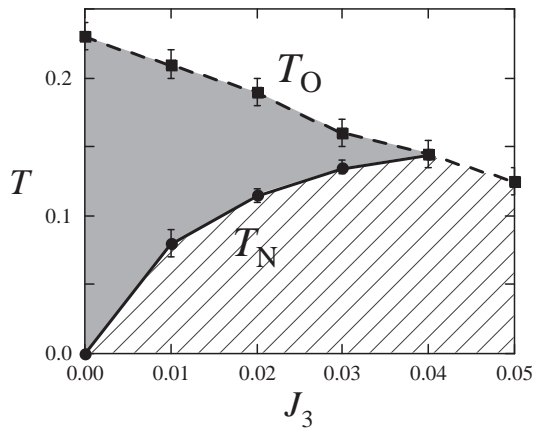}
\caption{
Phase diagram for the spin-orbital-lattice coupled model (\ref{eq:H}) 
determined by classical Monte Carlo calculations. 
The shaded area $T_{\rm N} < T < T_{\rm O}$ shows 
the orbital and lattice ordered phase. 
The hatched area below $T_{\rm N}$ is the AF ordered phase 
concomitant with the orbital and lattice orders. 
The orbital and lattice transition at $T_{\rm O}$ (the dashed line) 
is a first-order transition, 
while the AF transition at $T_{\rm N}$ (the solid line) is 
a second-order one. The lines are guides for the eyes. 
} 
\label{fig:phase diagram} 
\end{figure}

Figure~\ref{fig:phase diagram} summarizes the phase diagram 
in the $J_3$-$T$ parameter space determined by the analysis of the results in 
Figs.~\ref{fig:J_3 dep of C}-\ref{fig:J_3 dep of M_O and M_S} 
as well as other quantities such as $E$, $\bar{Q}$, $\chi_{\rm S}$, and $g_{\rm S}$. 
The shaded area for $T_{\rm N} < T < T_{\rm O}$ shows 
the orbital ordered phase with the tetragonal JT distortion 
with a flattening of VO$_6$ octahedra as shown in Fig.~\ref{fig:orbital order} (a). 
The 1D anisotropy of spin correlation found in Sec.~\ref{Sec:1D} 
appears in this region. 
The hatched area for $T < T_{\rm N}$ is the AF ordered phase 
concomitant with the orbital and lattice orders. 

From estimates of parameters in Sec.~\ref{Sec:Parameters}, 
we obtained $J \sim 200$K and $J_3 \sim 0.02J \sim 4$K. 
From Fig.~\ref{fig:phase diagram}, 
these estimates give $T_{\rm O} \sim 40$K and $T_{\rm N} \sim 20$K. 
The experimental values are $T_{\rm c1} \simeq 50$K and 
$T_{\rm c2} \simeq 40$K in ZnV$_2$O$_4$. 
\cite{Ueda1997} 
Thus, two transitions at $T_{\rm c1}$ and $T_{\rm c2}$ in experiments 
are consistently understood 
by the orbital and lattice ordering transition at $T_{\rm O}$ and 
the AF ordering transition at $T_{\rm N}$, respectively. 
The semiquantitative agreement of the transition temperatures 
between our MC results and the experimental values is satisfactory 
with considering the assumptions on the derivation of the model (\ref{eq:H}) and 
on parameter estimates in Sec.~\ref{Sec:Parameters}. 
In particular, we note that $\gamma$ and $\lambda$ in the JT part in Eq.~(\ref{eq:H_JT}), 
are parameters in our theory for which there has not been any experimental 
estimate to our knowledge as mentioned in Sec.~\ref{Sec:Parameters}. 
The agreement in spite of these assumptions strongly suggests that  
our model (\ref{eq:H}) captures essential physics 
in vanadium spinel oxides $A$V$_2$O$_4$.

\section{Discussions}
\label{Sec:Discussions}

\subsection{Role of tetragonal JT distortion}
\label{Sec:JTrole}

The orbital and lattice orderings at $T_{\rm O}$ are considered to 
be caused by the cooperation between the intersite orbital interaction 
in Eq.~(\ref{eq:H_SO}) and the JT coupling 
in Eq.~(\ref{eq:H_JT}). 
In this section, we examine the role of the JT distortion 
in more detail. 

We focus on the instability in the orbital sector and 
neglect spins momentarily. 
That is, we here consider the effective orbital model 
which is derived from the model (\ref{eq:H_SO}) 
by assuming the spin paramagnetic state in the mean-field level 
as discussed in Ref.~\onlinecite{Tsunetsugu2003}. 
We replace $\mib{S}_i \cdot \mib{S}_j$ by 
$\langle \mib{S}_i \cdot \mib{S}_j \rangle = 0$ in Eq.~(\ref{eq:H_SO}), and obtain  
the effective orbital Hamiltonian in the form 
\begin{eqnarray}
H_{\rm O} = J_{\rm O} \sum_{\langle i,j \rangle} 
n_{i \alpha(ij)} n_{j \alpha(ij)}, 
\label{eq:H_O}
\end{eqnarray}
where $J_{\rm O} = J (2A-C)$.
For simplicity, we neglect small contributions from the $J_3$ terms in this section. 
Since $J_{\rm O} > 0$, the Hamiltonian is an AF three-state clock model, 
which is the same as Eq.~(5) in Ref.~\onlinecite{Tsunetsugu2003} 
up to irrelevant constants. 

The mean-field argument for the model (\ref{eq:H_O}) predicts 
that a degeneracy remains partially in the tetrahedron unit. 
\cite{Tsunetsugu2003} 
There are totally $3^4 = 81$ different orbital states 
in the four-sites unit, and 
$30$ states among them are in the lowest energy state 
which have four antiferro-type and two ferro-type orbital bonds. 
The $30$ degenerate states are categorized into two different types 
shown in Fig.~2 (a) and (b) in Ref.~\onlinecite{Tsunetsugu2003}. 
[The former corresponds to Fig.~\ref{fig:orbital order} (b) and 
the latter corresponds to Figs.~\ref{fig:orbital order 2} (a) and (b) 
in the present paper.] 
In the first type, two ferro-type bonds do not touch with each other, 
while they touch at one site in the second type. 
Thus, we expect that in the absence of the JT coupling, 
the orbital degeneracy remains partially and 
prevents the emergence of a particular ordered state. 

\begin{figure}
\includegraphics[width=7cm]{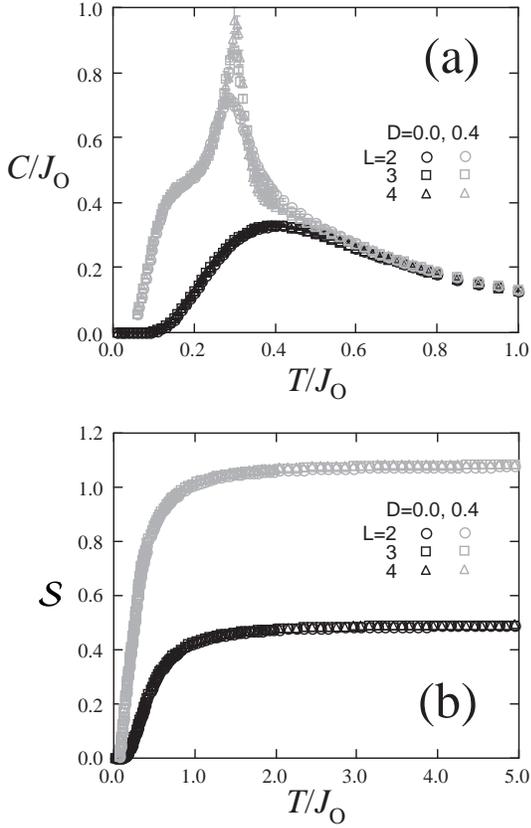}
\caption{
Temperature dependences of (a) the specific heat per site and 
(b) the entropy sum [Eq.~(\ref{eq:entropy sum})] 
for the effective orbital model (\ref{eq:H_O}). 
Black symbols show the results without the JT distortion. 
Gray symbols show the results in the presence of 
the tetragonal JT level splitting with $D/J_{\rm O} = 0.4$ 
in Eq.~(\ref{eq:H_D}).
} 
\label{fig:C and S} 
\end{figure}

To confirm this prediction, we perform the Monte Carlo simulation 
for the effective orbital model (\ref{eq:H_O}). 
The result of the specific heat per site 
is shown in Fig.~\ref{fig:C and S} (a) (black symbols). 
The data show a broad peak around $T \simeq 0.4J_{\rm O}$ 
without any singularity or 
any significant system-size dependence. 
This indicates that the system does not show any phase transition, 
as predicted by the above mean-field picture. 
In Fig.~\ref{fig:C and S} (b) (black symbols), 
we plot the temperature dependence of 
the entropy sum defined by 
\begin{equation}
{\cal S}(T) = \int_0^T \frac{C(T')}{T'} dT'. 
\label{eq:entropy sum}
\end{equation}
The entropy sum appears to saturate at $\sim 0.5$ at high temperatures, 
which is largely suppressed from the expected value of $\ln 3 \simeq 1.1$ 
for free three-state clock spins. 
This indicates that there remains the entropy at zero temperature 
due to the degeneracy discussed above. 

One way to estimate the zero-temperature entropy ${\cal S}_0$ is 
the so-called Pauling's method. 
\cite{Liebmann1986} 
In this method, the ground state degeneracy in the whole system 
is calculated simply multiplying the local degeneracy. 
In the present case, the total number of states is $3^{N_{\rm site}}$, and 
$30$ of $81$ states constitute 
the degenerate ground-state in each tetrahedron as mentioned above. 
With noting that there are $N_{\rm site}/2$ tetrahedra in the system and 
assuming that the tetrahedra are independent with each other, 
the zero-temperature entropy within the Pauling's approximation is obtained as
\begin{eqnarray}
{\cal S}_0 &=& \lim_{N_{\rm site} \to \infty} \frac{1}{N_{\rm site}} \ln \Big[ 
3^{N_{\rm site}} \Big( \frac{30}{81} \Big)^{N_{\rm site}/2} \Big]
\nonumber \\
&=& \ln 3 - \frac12 \ln \Big( \frac{81}{30} \Big). 
\end{eqnarray}
Thus, we obtain the entropy sum as 
\begin{equation}
{\cal S}(T \to \infty) = \ln 3 - {\cal S}_0 = \frac12 \ln \Big( \frac{81}{30} \Big) 
\simeq 0.5.
\end{equation}
The estimate is close to the saturated value in Fig.~\ref{fig:C and S} (b). 

The tetragonal JT distortion with a flattening of the octahedra 
as shown in Fig.~\ref{fig:orbital order} (a) 
splits the degenerate energy levels, and 
lowers the energy of one orbital state of the first type, i.e., 
the configuration in Fig.~\ref{fig:orbital order} (b). 
In this case, all the tetrahedra are in the same orbital states. 
That is, there we expect the orbital ordering and disappearance of 
the zero-temperature entropy. 
The results for the case with the level splitting are also shown 
in Fig.~\ref{fig:C and S} (gray symbols). 
Here, for simplicity, we incorporate the level splitting by adding the term 
\begin{equation}
H_D = -\frac{D}{3} \sum_i (n_{i1} + n_{i2} - 2 n_{i3}) 
\label{eq:H_D}
\end{equation}
to Eq.~(\ref{eq:H_O}), 
which mimics the JT term in Eq.~(\ref{eq:H_JT}). 
As expected, the specific heat in Fig.~\ref{fig:C and S} (a) 
shows a singularity at $T \sim 0.3 J_{\rm O}$, 
which indicates a phase transition. 
Figure~\ref{fig:C and S} (b) shows ${\cal S}(T \to \infty) \simeq \ln 3$, 
which indicates that the phase transition reduces the remaining entropy 
by lifting the degeneracy. 

The results in this section confirm the mean-field picture 
in Ref.~\onlinecite{Tsunetsugu2003}, and 
explicitly show the importance of the cooperation 
between the intersite orbital interaction and the JT coupling.

\subsection{Symmetry of orbital ordered state}
\label{Sec:symmetry}

Here, we comment on the spatial symmetry of 
the orbital ordered state below $T_{\rm O}$. 
Our result in Fig.~\ref{fig:orbital structure} indicates that 
the orbital ordering breaks the mirror symmetry 
for the $[110]$ or $[1\bar{1}0]$ plane. 
The symmetry breaking will also appear in the lattice structure 
through the electron-phonon coupling 
which breaks the $d_{yz}$ and $d_{zx}$ symmetry 
although such coupling is not included in our model 
[the tetragonal JT mode in Eq.~(\ref{eq:H_JT}) 
does not split the $d_{yz}$ and $d_{zx}$ levels]. 

Recently, the orbital ordered state with another symmetry 
has been proposed theoretically. 
\cite{TchernyshyovPREPRINT} 
The theory is based on the assumption of the dominant role of 
the relativistic spin-orbit coupling. 
The predicted orbital order consists of 
the ferro-type occupation of 
the $d_{xy}$ and $(d_{yz} + i d_{zx})$ orbitals at every site. 
This orbital order does not break the mirror symmetry. 

The difference of the resultant orbital ordered state is 
important to consider the fundamental physics 
of the present $t_{2g}$ electron system. 
In $t_{2g}$ electron systems, in general, 
there is keen competition among different contributions 
whose energy scales are close to each other; 
the spin and orbital exchange interactions, 
the JT coupling, 
and the relativistic spin-orbit coupling. 
\cite{Kugel1982}
In our argument, the former two mechanisms play a primary role 
while the last one is not taken into account. 
On the contrary, in Ref.~\onlinecite{TchernyshyovPREPRINT}, 
the relativistic spin-orbit coupling is assumed to be dominant. 
Therefore, the determination of the orbital and lattice symmetry 
is relevant to clarify the primary factor in the keen competition. 

Experimental results on the lattice symmetry, however, 
are still controversial. 
In Refs.~\onlinecite{Nishiguchi2002} and \onlinecite{Reehuis2003}, 
the X-ray scattering results for polycrystal samples 
are consistent with the symmetry $I4_1/amd$ below $T_{\rm c1}$, 
which suggests the persistence of the mirror symmetry 
at the low temperature phase. 
On the contrary, in recent synchrotron X-ray data 
for a single crystal sample, a new peak is found 
whose intensity is three orders of magnitude weaker than a typical main peak. 
\cite{LeePREPRINT} 
Unfortunately, the new peak cannot conclude 
whether the mirror symmetry is broken or not, but 
it suggests the different symmetry from the previous $I4_1/amd$, 
i.e., either $I\bar{4}m2$ or $I\bar{4}$. 
This controversy, probably coming from a small electron-phonon coupling 
in this $t_{2g}$ system, 
reveals the necessary of more sophisticated experiments 
for larger single crystals. 
Such experiments are highly desired to settle the theoretical controversy. 

Another possible experiment to conclude the symmetry problem 
is to detect the orbital state directly. 
This may be achieved, for instance, 
by the resonant X-ray scattering technique. 
\cite{Murakami1998} 
Besides, another way to distinguish the orbital ordered states 
in our results and in Ref.~\onlinecite{TchernyshyovPREPRINT} 
is to examine the time reversal symmetry. 
The $(d_{yz} + i d_{zx})$ orbital order 
in Ref.~\onlinecite{TchernyshyovPREPRINT} 
breaks the time reversal symmetry 
even above the magnetic transition temperature. 
A possible experiment is the detection of 
either circular dichroism or birefringence.

\subsection{Quantum fluctuation effect}
\label{Sec:quantum fluctuation}

Our classical Monte Carlo calculations have 
shown that with approaching zero temperature, 
the staggered moment increases and saturates 
at the maximum value, $M_{\rm S} \rightarrow 1$, 
when the third-neighbor coupling $J_3$ is finite. 
In physical units, this value should be 
multiplied by $gS \mu_{\rm B}$ and 
corresponds to $M_{\rm S} = 2 \mu_{\rm B}$. Here, 
$\mu_{\rm B}$ is the Bohr magneton and  
the $g$-factor is assumed to be the standard 
value $g=2$.  
This behavior was plotted in Figs.~\ref{fig:spin} (a) 
and \ref{fig:J_3 dep of M_O and M_S} (b)-(f). 
On the other hand, recent experiments show 
$M_{\rm S} \sim 0.6 \mu_{\rm B}$ at low temperatures, 
only less than a half of this classical value. 
\cite{LeePREPRINT,Reehuis2003}
It is well known that quantum fluctuations 
due to magnon excitations reduce the amplitude 
of spontaneous moment in antiferromagnets. 
We expect that this effect is particularly 
important in frustrated magnets like pyrochlore 
systems, since frustration generally reduces 
the energy scale of magnon excitations 
and correspondingly enhances quantum fluctuations 
of the staggered moment.  

Here, we examine at $T=0$ the effect of quantum 
fluctuations of spin degree of freedom 
on the reduction of staggered moment 
for the model (\ref{eq:H}), by using the 
linear spin wave theory. For this purpose, 
we assume the perfect orbital order 
with the ordering pattern in Fig.~\ref{fig:orbital structure}, 
and substitute the density operators in Eq.~(\ref{eq:H_SO}) 
by their expectation values 0 or 1. 
That is, we here consider 
the effective quantum spin model on the pyrochlore 
lattice in the form 
\begin{equation}
H_{\rm spin} = 
\sum_{\langle i,j \rangle} 
J_{ij} \mib{S}_i \cdot \mib{S}_j
+ {\sum_{\langle\!\langle i,j \rangle\!\rangle}} 
J_{3}' \mib{S}_i \cdot \mib{S}_j,
\label{sw_eq1}
\end{equation}
where the first summation should be taken over 
nearest-neighbor pairs, 
while the second one be over third-neighbor 
pairs coupled by $\sigma$-bond.  
Under the orbital ordering in Fig.~\ref{fig:orbital structure}, 
exchange coupling constants are 
ferromagnetic for nearest-neighbor spin pairs 
on the $yz$ and $zx$ chains, 
$J_{ij}= -J B =J_{\rm F} < 0$,   
and antiferromagnetic for those on the $xy$ chains, 
$J_{ij}= J C =J_{\rm AF} >0$, 
while the third-neighbor couplings are 
also antiferromagnetic, $J_{3}' = J_3 C$.  
These parameters were defined in Eqs.~(\ref{eq:J})-(\ref{eq:eta}).  
It is important that 
$J_{\rm AF} \gg |J_{\rm F}| \gg J_{3}'$, 
since $\eta \ll 1$ and 
$t_\sigma^{\rm 3rd} \ll t_\sigma^{\rm nn}$.  
Roughly speaking, the small control parameter 
in the spin wave theory is 
1/[$S\times z$ (number of neighbors)]. 
In the present case, $S$=1 and $z$=6, leading to 
a control parameter small enough, and therefore we may 
expect quite good estimate from the spin wave theory.  

For the magnetic order determined by the mean 
field arguments in Sec.~\ref{Sec:Spin-Orbital Model} and 
Ref.~\onlinecite{Tsunetsugu2003}, we have calculated 
the reduction of staggered moment at $T=0$ by using 
the linear spin wave theory.  
The magnetic structure is 
shown in Fig.~\ref{fig:spin order} and its unit cell contains 8 
spins.  Details of the calculations will be 
reported elsewhere and here we show only the 
results for the staggered moment.  

When the third-neighbor exchange couplings are 
zero ($J_{3}'=0$), magnons have zero energy (zero modes) 
on the planes $k_x = \pm k_y$
in the Brillouin zone.  
Within the linear spin wave theory, magnons 
are treated as noninteracting to each other, 
and these zero modes are excited freely without 
any energy cost.  This leads to the logarithmic divergence 
of the reduction of moment, 
$\Delta S \rightarrow \infty$.  
It is noted that in the pyrochlore spin system 
in which all the nearest-neighbor couplings 
are antiferromagnetic with same amplitude, 
a half of magnon excitations are zero modes throughout 
the whole Brillouin zone, 
and this results in $\Delta S = \infty$, 
as a manifestation of strong geometrical frustration.
\cite{Tsunetsugu2002a}

\begin{figure}
\includegraphics[width=7cm]{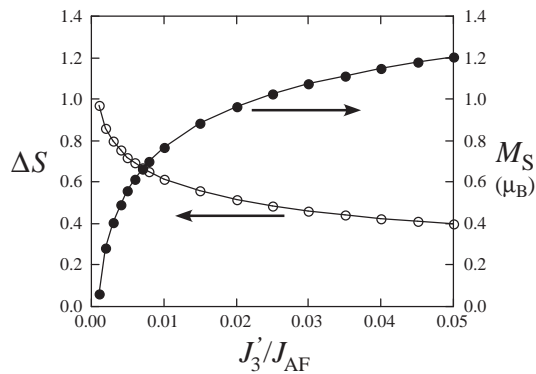}
\caption{
Reduction of the staggered moment due to quantum fluctuations (open symbols) 
estimated by the linear spin wave calculation 
at $J_{\rm F}/J_{\rm AF} = -0.113$. 
The staggered moment is also plotted (filled symbols), 
which is given by $M_{\rm S} = (S - \Delta S) g \mu_{\rm B}$ 
and by assuming $g=2$. 
The lines are guides for the eyes. 
} 
\label{fig:deltaS} 
\end{figure}

Once the third-neighbor exchange couplings $J_{3}'$
are switched on, zero modes acquire a positive 
energy and the reduction of moment becomes finite.  
Figure~\ref{fig:deltaS} shows the results of the reduction 
of moment $\Delta S$ as a function of $J_{3}'$ 
for $J_{\rm AF}=1$ and $J_{\rm F}=-0.113$, 
corresponding to $\eta = 0.08$ at which the MC calculations 
have been performed. 
The amplitude of staggered moment is reduced to 
$M_{\rm S} = (S - \Delta S) g \mu_{\rm B}$. 
In the case of the vanadium spinel oxides with $S=1$, 
$S - \Delta S > 0$ for $J_{3}' \simge 10^{-3} J_{\rm AF}$, which indicates that 
the antiferromagnetic order shown in Fig.~\ref{fig:spin order} 
is stable down to very small third-neighbor 
exchange couplings.  
For example, at a reasonable value, 
$J_{3}' \sim 0.02 J_{\rm AF}$, 
the amplitude of staggered moment is 
$M_{\rm S} \sim (1 - 0.5)g\mu_{\rm B} \sim 0.5 g\mu_{\rm B}$.  With using 
$g=2$, this corresponds to 
$M_{\rm S} \sim 1 \mu_{\rm B}$.  
Hence, the reduction is significantly large 
in the present spin-orbital-lattice coupled system, and 
particularly the staggered moment rapidly decreases 
as $J_{3}'$ in the realistic small-$J_{3}'$ region. 
We consider that 
the estimate of $M_{\rm S}$ is satisfactory 
and can explain the experimental value 
by carefully tuning the model parameters. 

We also note that there may be a small contribution from 
quantum fluctuations in the orbital degree of freedom. 
In our spin-orbital-lattice coupled model (\ref{eq:H}), 
this contribution is not taken into account 
since we consider only $\sigma$ bond for hopping integrals. 
Other hopping integrals lead to 
small orbital exchange interactions, 
causing small quantum fluctuations of orbitals. 
This will slightly reduce the orbital sublattice moment and 
may lead to a further reduction of the staggered spin moment.

\subsection{Difference of ZFC and FC susceptibility}
\label{Sec:ZFC and FC}

We have shown that the MC results for the uniform magnetic susceptibility 
in Sec.~\ref{Sec:chi} qualitatively explain 
the experimental data in the zero-field-cool (ZFC) measurement. 
In experiments, there has been reported that 
the ZFC and field-cool (FC) data show 
a large difference at low temperatures. 
\cite{Ueda1997} 
The difference between the ZFC and FC data emerges 
at a higher temperature than the structural transition temperature $T_{\rm c1}$,  
and develops on cooling. 
One of the mysteries is that the difference remains finite and large 
even in the AF spin ordered phase below $T_{\rm c2}$: 
The FC value becomes nearly twice of the ZFC one at the lowest temperature. 
This strongly suggests that the difference cannot be explained as 
a simple spin-glass phenomenon. 

A key observation is that the starting temperature of the difference 
between the ZFC and FC data 
as well as the magnitude of the difference 
appears to depend on samples. 
In Ref.~\onlinecite{Ueda1997}, the difference emerges at $T \sim 100$K, 
and becomes $\sim 0.8 \times 10^3$emu/V-mol at $T_{\rm c1}$. 
On the other hand, in Ref.~\onlinecite{Reehuis2003}, 
the difference starts at $T \sim 70$K, and 
becomes $\sim 0.3 \times 10^3$emu/V-mol at $T_{\rm c1}$. 
We also note that the temperature dependences of the ZFC data 
show a sample dependence, especially near two transition temperatures. 
These sample dependences imply the importance of quenched disorder. 
Effects of disorder in the present spin-orbital-lattice coupled system 
are interesting and open problems. 

In our MC calculations, it is difficult to obtain the FC results 
because the system shows a discontinuous transition at $T_{\rm O}$: 
There appears a huge hysteresis when we change the temperature 
successively by using the MC sample in the previous run 
as the initial state in the next run. 
On cooling, the system remains in the high-temperature para state 
well below the true transition temperature $T_{\rm O}$ 
as a metastable state because of an exponentially large energy barrier 
in the first-order transition. 
One way to avoid this hysteresis is to apply 
more sophisticated MC method such as the multicanonical technique. 
\cite{Berg1991}
This interesting problem is left for further study.

\subsection{$A$-site substitution effect}
\label{Sec:A-site}

We have shown that our spin-orbital-lattice coupled model (\ref{eq:H}) 
exhibits two transitions which well agree with 
experimental results in $A$V$_2$O$_4$. 
Here, we discuss a difference among the compounds 
with different $A$ cations observed in experiments. 

Among $A$V$_2$O$_4$ with $A$=Zn, Mg, or Cd, 
CdV$_2$O$_4$ shows a different behavior near $T_{\rm c1}$ from others. 
\cite{Nishiguchi2002} 
As temperature decreases, 
the magnetic susceptibility for the compounds with $A$=Zn and Mg 
shows a sudden drop at $T_{\rm c1}$ 
as in our numerical result in Fig.~\ref{fig:chi}, 
whereas that for the Cd compound shows a sharp increase. 
Moreover, the transition temperature $T_{\rm c1}$ is 
substantially higher in the Cd case than in the Zn and Mg cases; 
$T_{\rm c1} = 97$K for Cd 
while $T_{\rm c1} = 52$K and $64.5$K for Zn and Mg, respectively. 
On the other hand, 
the AF magnetic transition temperature $T_{\rm c2}$ is 
almost the same among three compounds; 
$T_{\rm c2} = 44, 45$, and $35$K for Zn, Mg, and Cd, respectively. 
\cite{Nishiguchi2002} 

Our results show that the transition temperature 
$T_{\rm O}$, which corresponds to $T_{\rm c1}$, 
is determined by the energy balance between 
the intersite orbital interaction in Eq.~(\ref{eq:H_SO}) and 
the elastic energy gain in Eq.~(\ref{eq:H_JT}). 
The $A$-site cations are nonmagnetic and locate at the tetrahedral sites 
which are separated from the pyrochlore network of V cations, and  
the change of the $A$ cation may affect the lattice structure 
since $A$O$_4$ tetrahedra share oxygen ions with VO$_6$ octahedra. 
Thereby, we may consider that 
the Cd cation, which has relatively large ionic radius, 
may modify the lattice structure 
and have an influence on 
the JT energy gain as well as the form and magnitude of 
the orbital interaction. 

Actually, although the symmetry is the same for the three compounds, 
the so-called $u$ parameter of the spinel structure is considerably larger 
for CdV$_2$O$_4$ than for Zn and Mg compounds. 
The $u$ parameter represents the magnitude of the trigonal distortion. 
The absence of the trigonal distortion gives $u=0.375$ by definition, 
and a larger $u$ denotes a larger trigonal distortion. 
For the Zn and Mg cases, $u = 0.385$ and $0.386$, respectively, 
whereas $u$ becomes 0.394 for the Cd compound: 
The trigonal distortion is substantially larger in the Cd case 
than in the Zn and Mg cases. 
This suggests that the peculiar behavior in CdV$_2$O$_4$ may be 
ascribed to effects of the trigonal distortion. 

The agreement between the experimental data in Zn and Mg compounds 
and our numerical results, 
which do not include effects of the trigonal distortion, 
indicates that the small trigonal distortion does not play 
an important role in these two compounds. 
To understand the behavior of the Cd compound, i.e., 
the sharp increase in the magnetic susceptibility 
as well as the higher $T_{\rm c1}$, 
it is interesting to examine the effect of 
the trigonal distortion in our theoretical framework. 
In addition to the contribution from the electron-phonon coupling 
to the trigonal distortion, 
we may have to include more complicated hopping integrals 
not only the $\sigma$-bond type but also, for instance, 
$\pi$-bond type or the second-neighbor hoppings, 
since the trigonal distortion affects 
the network of VO$_6$ octahedra: 
These additional hoppings modify the spin and orbital intersite interactions 
in the derived effective model in Eq.~(\ref{eq:H_SO}). 
The complicated effects of this trigonal distortion are 
left for further study.

\section{Summary}
\label{Sec:Summary}

In the present work, we have investigated the microscopic mechanism 
of two transitions in vanadium spinel oxides 
$A$V$_2$O$_4$ with nonmagnetic divalent $A$ cations such as Zn, Mg, and Cd. 
We have focused on the role of the $t_{2g}$ orbital degree of freedom 
as well as spin in these strongly correlated electron systems 
on the geometrically frustrated lattice structure, 
i.e., the pyrochlore lattice consists of the magnetic V cations. 
We have derived the effective spin-orbital-lattice coupled model 
in the strong correlation limit from the multiorbital Hubbard model 
with explicitly taking account of the $t_{2g}$ orbital degeneracy. 
The effective model describes the interplay 
between orbital and spin, and 
reveals the contrasting form of the intersite interactions 
in two degrees of freedom; 
the Heisenberg type for the spin part and 
the three-state clock type for the orbital part. 
The anisotropy in the orbital interaction 
originates from the dominant role of the $\sigma$-bond hopping integrals 
in the edge-sharing configuration of VO$_6$ octahedra. 
The Jahn-Teller coupling with the tetragonal lattice distortion 
is also included. 
Thermodynamic properties of the effective model have been investigated 
by the Monte Carlo simulation, which is 
a classical one to avoid the negative sign problem 
due to the geometrical frustration. 
Quantum corrections are examined by the spin wave approximation. 
Main results are summarized in the following. 

Our effective spin-orbital-lattice coupled model exhibits two transitions 
at low temperatures. 
As temperature decreases, first, the orbital ordering transition occurs 
assisted by the tetragonal Jahn-Teller distortion. 
This is a first-order transition. 
Successively, at a lower temperature, 
the antiferromagnetic spin order sets in. 
This transition is second order. 

For realistic parameter values, our numerical results agree 
with experimental data semiquantitatively. 
The estimates of two transition temperatures are 
$T_{\rm O} \sim 40$K and $T_{\rm N} \sim 20$K, 
while the experimental values are $T_{\rm c1} \sim 50$K and $T_{\rm c2} \sim 40$K. 
The changes of the entropy at the transitions are 
comparable to the experimental values: 
In particular, the small change at $T_{\rm N}$ 
compared to the considerable change at $T_{\rm O}$ is well reproduced. 
The magnetic ordering structure in the low temperature phase 
is completely consistent with the neutron scattering results. 
The magnitude of the staggered moment at $T=0$ 
is largely reduced by quantum fluctuations, and 
the estimate by the spin wave theory reasonably 
agrees with the experimental data. 
The temperature dependence of the uniform magnetic susceptibility 
is similar to the experimental data. 
The agreement strongly indicates that 
our effective model captures the essential physics of 
the vanadium spinel compounds $A$V$_2$O$_4$. 

Our results give an understanding of the mechanism of the two transitions 
in the vanadium spinel oxides.
The present numerical study has confirmed the mean-field scenario 
in our previous publication, 
\cite{Tsunetsugu2003}
and moreover, given more detailed and quantitative information. 
The first transition with orbital and lattice orderings 
is induced by the intersite orbital interaction 
which is three-state clock type and 
spatially anisotropic depending on both the bond direction and 
the orbital states in two sites. 
The anisotropy lifts the degeneracy due to the geometrical frustration 
inherent to the pyrochlore lattice. 
The tetragonal Jahn-Teller coupling also plays an important role 
to stabilize the particular orbital ordering pattern. 
The obtained orbital ordering structure is the antiferro type 
which consists of the alternative stacking 
of two ferro-type $ab$ planes; 
($d_{xy}, d_{yz}$) orbitals are occupied in one plane, and 
($d_{xy}, d_{zx}$) orbitals are occupied in the other.
Once the orbital ordering takes place, 
the orbital state affects the spin exchange interactions 
through the spin-orbital interplay, and 
reduces the magnetic frustration partially. 
As a consequence, the antiferromagnetic spin correlation develops 
mainly in the one-dimensional chains in the $ab$ planes. 
We found typically about three-times longer correlation length 
in the $xy$ direction than in the $yz$ and $zx$ directions. 
At the second transition, these one-dimensional chains 
are ordered by the third-neighbor exchange interaction 
to form the three-dimensional antiferromagnetic spin order. 
It was found that thermal fluctuations stabilize 
the collinear state by the order-by-disorder type mechanism. 

There still remain several open problems on this topic 
as discussed in Sec.~\ref{Sec:Discussions}; 
the controversy on the symmetry of the orbital ordered state, 
the large difference of the zero-field-cool and field-cool data 
of the magnetic susceptibility, and 
the $A$-site dependence, 
which is probably related to the trigonal distortion. 
They need further investigations from both experimental and 
theoretical viewpoints. 
Besides them, another important problem is 
the carrier doping effect on the Mott insulating materials, 
such as the Li doping in Zn$_{1-x}$Li$_x$V$_2$O$_4$. 
In experiments, it is known that the doping rapidly 
destroys the ordered states at $x=0$, and 
replaces them by a glassy state. 
\cite{Urano2000}
Finally, at $x=1$, the system becomes metallic and shows 
a heavy-fermion like behavior. 
\cite{Kondo1997,Urano2000} 
We believe that our present results give a good starting point 
to study the interesting doping effects.

\section*{Acknowledgment}

This work was supported by a Grant-in-Aid and NAREGI Nanoscience Project 
from the Ministry of Education, Science, Sports, and Culture. 
A part of the work was accomplished during Y. M. was staying 
at the Yukawa Institute of Theoretical Physics, Kyoto University, 
with the support from The 21st Century for Center of Excellence program, 
`Center for Diversity and Universality in Physics'.

\section*{Appendix A}

In this Appendix, we examine the conditions 
for $\gamma$ and $\lambda$ in the Hamiltonian (\ref{eq:H_JT}) 
to stabilize the orbital order in Fig.~\ref{fig:orbital order} (b) 
accompanied by the tetragonal JT distortion 
with a flattening of VO$_6$ octahedra 
as shown in Fig.~\ref{fig:orbital order} (a) at low temperatures. 
We employ the mean-field type argument to discuss the instability 
at high temperatures as in Ref.~\onlinecite{Tsunetsugu2003}. 
Here, we consider only the nearest-neighbor interactions and 
neglect small contributions from $H_{\rm SO}^{\rm 3rd}$. 
With considering a spin disordered state and replacing 
$\mib{S}_i \cdot \mib{S}_j$ by 
$\langle \mib{S}_i \cdot \mib{S}_j \rangle = 0$ in Eq.~(\ref{eq:H}), 
we obtain the effective Hamiltonian for the orbital and JT parts as 
\begin{equation}
H_{\rm O-JT} = 
J_{\rm O} \sum_{\langle i,j \rangle} n_{i \alpha(ij)} n_{j \alpha(ij)} 
+ H_{\rm JT}, 
\label{eq:H_orb-JT} 
\end{equation}
where $J_{\rm O} = J (2A-C)$. 
The first term is equivalent to Eq.~(\ref{eq:H_O}) and 
to Eq.~(5) in Ref.~\onlinecite{Tsunetsugu2003} up to irrelevant constants. 

We consider the orbital state and 
the JT distortion in the tetrahedron unit  
shown in Fig.~\ref{fig:orbital order} (b). 
For an orbital configuration at the four sites, 
the expectation value of the JT part $H_{\rm JT}$ for one tetrahedron 
is written in the following quadratic form 
\begin{equation}
e_{\rm JT}(\mib{Q}) 
= \mib{Q} \cdot \hat{M} \mib{Q} + \mib{Q} \cdot \mib{A} 
+ \mib{A} \cdot \mib{Q}, 
\label{eq:e_JT}
\end{equation}
where $\mib{Q}$ is a four-dimensional vector which denotes 
the amplitude of the JT distortion in each site as 
$\mib{Q} = ^t(Q_1,Q_2,Q_3,Q_4)$, and 
$\hat{M}$ is the $4 \times 4$ matrix whose matrix elements are given by 
$\hat{M}_{ij} = 1/2$ for $i=j$ and $\hat{M}_{ij} = -\lambda$ for $i \neq j$. 
The vector $\mib{A}$ describes the electron-phonon coupling part 
and depends on the orbital state; 
for instance, $\mib{A} = ^t(-\gamma/2,-\gamma/2,-\gamma/2,-\gamma/2)$ 
for the orbital occupation in Fig.~\ref{fig:orbital order} (b). 
Here, we consider the physical situation in which 
the system does not show any spontaneous lattice distortion 
without the coupling to electrons 
of the second and third terms in Eq.~(\ref{eq:e_JT}). 
This is satisfied when the matrix $\hat{M}$ is positive definite. 
Three of the eigenvalues of $\hat{M}$ are $(1+2\lambda)/2$ and 
the last one is $(1-6\lambda)/2$, 
and therefore, by noting that we consider only positive $\lambda$ 
as mentioned in Sec.~\ref{Sec:Spin-Orbital Model}, 
the condition is 
\begin{equation}
0 < \lambda < \frac16. 
\label{eq:cond}
\end{equation}

The quadratic form of $e_{\rm JT}$ in Eq.~(\ref{eq:e_JT}) 
can be transformed to the following form 
\begin{equation}
e_{\rm JT}(\mib{P}) = (\mib{P} + \mib{B}) \cdot (\mib{P} + \mib{B}) 
- \mib{B} \cdot \mib{B}, 
\end{equation}
where $\mib{P} = \hat{L} \mib{Q}$, $\mib{B} = \hat{L}^{-1} \mib{A}$, and 
$\hat{L} = \hat{M}^{1/2}$ is 
well-defined since $\hat{M}$ is a positive-definite matrix 
under the condition of Eq.~(\ref{eq:cond}). 
Therefore, we find that the JT energy takes its minimum value 
\begin{equation}
e_{\rm JT}^{\rm min} = - \mib{B} \cdot \mib{B} 
= - \mib{A} \cdot \hat{M}^{-1} \mib{A}, 
\label{eq:e_JT^min}
\end{equation}
for the amplitude of the distortions 
\begin{equation}
\mib{Q}^{\rm min} = - \hat{M}^{-1} \mib{A}. 
\end{equation}
For instance, for the configuration in Fig.~\ref{fig:orbital order} (b), 
we obtain $e_{\rm JT}^{\rm min} = -2 \gamma^2 / (1-6\lambda)$ 
for $\mib{Q}^{\rm min} = ^t(Q,Q,Q,Q)$ with $Q = \gamma/(1-6\lambda)$. 

All the different orbital configurations with 
$\langle n_{i\alpha} \rangle = 0$ or $1$ are classified 
by the energy value $e_{\rm JT}$ into five groups, 
which are described by the vector $\mib{A}$ as 
\begin{eqnarray}
\mib{A}_1 &=& ^t(-\gamma/2,-\gamma/2,-\gamma/2,\gamma), 
\label{eq:A_1} 
\\
\mib{A}_2 &=& ^t(-\gamma/2,-\gamma/2,\gamma,\gamma), 
\label{eq:A_2} 
\\
\mib{A}_3 &=& ^t(-\gamma/2,\gamma,\gamma,\gamma), 
\label{eq:A_3} 
\\
\mib{A}_4 &=& ^t(\gamma,\gamma,\gamma,\gamma), 
\label{eq:A_4} 
\\
\mib{A}_5 &=& ^t(-\gamma/2,-\gamma/2,-\gamma/2,-\gamma/2), 
\label{eq:A_5} 
\end{eqnarray}
where arbitrary permutations of components in each vector give the same energy. 
Typical orbital configurations are shown in Fig.~\ref{fig:orbital order 2}. 
(The configuration for $\mib{A}_5$ is shown in Fig.~\ref{fig:orbital order} (b).) 

\begin{figure}
\caption{
Orbital ordering patterns 
(a), (b), (c), and (d) for the categories 1, 2, 3, and 4 specified by 
Eqs.~(\ref{eq:A_1}), (\ref{eq:A_2}), (\ref{eq:A_3}), and (\ref{eq:A_4}), 
respectively. 
The ferro-type (antiferro-type) orbital bonds are shown 
by the blue solid (red dashed) lines. 
} 
\label{fig:orbital order 2} 
\end{figure}

On the basis of the above consideration on the JT energy, 
we calculate the expectation value of the Hamiltonian (\ref{eq:H_orb-JT}), 
for the five types of orbital configurations. 
With considering the orbital interaction energy 
from the first term in Eq.~(\ref{eq:H_orb-JT}), 
the minimized values of the energy per tetrahedron are obtained as 
\begin{eqnarray}
e_1 &=& 4J_{\rm O} - \frac{\gamma^2 (7-40\lambda)}{2 (1+2\lambda) (1-6\lambda)}, 
\\
e_2 &=& 4J_{\rm O} - \frac{\gamma^2 (5-26\lambda)}{(1+2\lambda) (1-6\lambda)}, 
\\
e_3 &=& 6J_{\rm O} - \frac{\gamma^2 (13-28\lambda)}{2 (1+2\lambda) (1-6\lambda)}, 
\\
e_4 &=& 8J_{\rm O} - \frac{8\gamma^2}{1-6\lambda}. 
\\
e_5 &=& 4J_{\rm O} - \frac{2\gamma^2}{1-6\lambda},
\end{eqnarray}
Note that there are twelve bonds per one tetrahedron. 

\begin{figure}
\includegraphics[width=6.5cm]{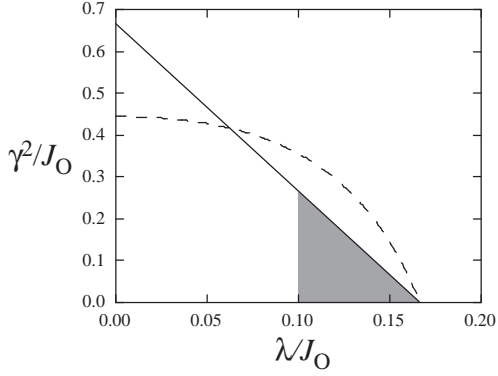}
\caption{
The parameter region of Eqs.~(\ref{eq:condition1})-(\ref{eq:condition3}) is 
shown by the shaded area.  
The dashed and solid lines denote the conditions of 
Eqs.~(\ref{eq:condition2}) and (\ref{eq:condition3}), respectively. 
} 
\label{fig:conditions} 
\end{figure}

By comparing $e_5$ with other four values $e_i$ ($i=1-4$), 
we obtain the stability conditions 
for the orbital configuration of Fig.~\ref{fig:orbital order} (b). 
First, the conditions $e_5 < e_1$ and $e_5 < e_2$ are satisfied when 
\begin{equation}
\frac{1}{10} < \lambda < \frac{1}{6}. 
\label{eq:condition1}
\end{equation}
In this range, the condition $e_5 < e_3$ leads to 
\begin{equation}
\gamma^2 < \frac{4J_{\rm O} (1+2\lambda) (1-6\lambda)}{9 (1-4\lambda)}, 
\label{eq:condition2}
\end{equation}
and the condition $e_5 < e_4$ leads to
\begin{equation}
\gamma^2 < 2 J_{\rm O} (1-6\lambda) / 3. 
\label{eq:condition3}
\end{equation}
The shaded area in Fig.~\ref{fig:conditions} shows the parameter region 
where all the conditions of Eqs.~(\ref{eq:condition1})-(\ref{eq:condition3}) 
are satisfied. 
We performed several MC runs to confirm the present mean-field type analysis. 
In the MC calculations in Sec.~\ref{Sec:Results}, 
we take $\gamma^2/J = 0.04$ and $\lambda/J = 0.15$ as a typical set in this region. 

\section*{Appendix B}

In this appendix, we derive the expressions for the spectral weights 
in Eqs.~(\ref{eq:I_x})-(\ref{eq:I_z}) for our effective model (\ref{eq:H}). 

The spectral weight, which is the total weight of the optical conductivity, 
is given by the kinetic energy of the system ($f$-sum rule). 
\cite{Maldague1977,Shastry1990}
We may safely extend the formula for the single-band Hubbard model; 
\begin{equation}
I_\mu = \frac{2}{\pi} \int_0^\infty \sigma_{\mu\mu}(\omega) d\omega 
= \frac{1}{N_{\rm u.c.}} \sum_{\mib{k}} \Big\langle - \frac{\partial^2 K(\mib{k})}
{\partial k_\mu^2} \Big\rangle,
\label{eq:I}
\end{equation}
to our multiorbital case. 
(We set $e = \hbar = 1$, and $N_{\rm u.c.}$ is 
the number of unit cells.) 
Here, $K(\mib{k})$ describes 
the kinetic energy term of the multiorbital Hubbard Hamiltonian as 
\begin{equation}
\sum_{\mib{k}} K(\mib{k}) = \sum_{\mib{k}} \sum_{a,b} \sum_{\alpha, \beta} 
t_{a \alpha, b \beta} (\mib{k}) c_{a \alpha}^\dagger (\mib{k}) 
c_{b \beta} (\mib{k}), 
\end{equation}
which is the Fourier transform of the first term of Eq.~(\ref{eq:H_multiorbital}). 
Here, $a,b = 0,1,2,3$ are the atom indices in the primitive unit-cell 
containing four V atoms, whose relative positions are labelled by 
$\mib{\delta}_0 = (0,0,0)$, $\mib{\delta}_1 = (0,1/4,1/4)$, 
$\mib{\delta}_2 = (1/4,0,1/4)$, $\mib{\delta}_3 = (1/4,1/4,0)$, respectively. 

Since the pyrochlore lattice consists of three different chains 
in the $xy$, $yz$, and $zx$ directions, and 
we here consider only the $\sigma$-bond hopping integrals 
which are finite only along the chains, 
$K(\mib{k})$ is given by the summation of 
the contributions from each chain as
\begin{equation}
K(\mib{k}) = \sum_{\zeta} \big[ 
u_{\zeta} \kappa_{\zeta}^{\rm nn} (\mib{k}) 
+ v_{\zeta} \kappa_{\zeta}^{\rm 3rd} (\mib{k}) 
\big], 
\label{eq:K}
\end{equation}
where $\zeta = 1$ ($yz$), $2$ ($zx$), $3$ ($xy$) denote 
the three different types of the chains, and 
$u_{\zeta}, v_{\zeta}$ are the nearest-neighbor and the third-neighbor 
hopping integrals along the $\zeta$ chain, respectively. 
In the Hamiltonian (\ref{eq:H_multiorbital}), 
$u_1 = u_2 = u_3 = t_{\sigma}^{\rm nn}$ and 
$v_1 = v_2 = v_3 = t_{\sigma}^{\rm 3rd}$, but 
they are deliberately denoted by different parameters for later use. 
Each term in Eq.~(\ref{eq:K}) is defined by 
\begin{eqnarray}
&& \kappa_{1}^{\rm nn} (\mib{k}) = 
2 \cos(\mib{k} \cdot \mib{\delta}_{01}) [ 
c_{01}^\dagger(\mib{k}) c_{11}(\mib{k}) + 
c_{11}^\dagger(\mib{k}) c_{01}(\mib{k}) ]
\nonumber \\
&& \quad + 
2 \cos(\mib{k} \cdot \mib{\delta}_{23}) [ 
c_{21}^\dagger(\mib{k}) c_{31}(\mib{k}) + 
c_{31}^\dagger(\mib{k}) c_{21}(\mib{k}) ],
\\
&& \kappa_{1}^{\rm 3rd} (\mib{k}) = 
2 \cos(2\mib{k} \cdot \mib{\delta}_{01}) [ 
c_{01}^\dagger(\mib{k}) c_{01}(\mib{k}) + 
c_{11}^\dagger(\mib{k}) c_{11}(\mib{k}) ]
\nonumber \\
&& \quad +  
2 \cos(2\mib{k} \cdot \mib{\delta}_{23}) [ 
c_{21}^\dagger(\mib{k}) c_{21}(\mib{k}) + 
c_{31}^\dagger(\mib{k}) c_{31}(\mib{k}) ],
\end{eqnarray}
and so on, where $\mib{\delta}_{ab} = \mib{\delta}_a - \mib{\delta}_b$.  

Let us now evaluate the derivatives of $K(\mib{k})$ in Eq.~(\ref{eq:I}). 
By using Eq.~(\ref{eq:K}), it is straightforward to obtain, 
for instance, the derivative in terms of $k_x$ in the form 
\begin{eqnarray}
\frac{\partial^2 K(\mib{k})}{\partial k_x^2} &=& 
\frac{1}{16} \big[ 
u_2 \kappa_2^{\rm nn}(\mib{k}) + u_3 \kappa_3^{\rm nn}(\mib{k}) \big]
\nonumber \\
&+& \frac14 \big[ 
v_2 \kappa_2^{\rm 3rd}(\mib{k}) + v_3 \kappa_3^{\rm 3rd}(\mib{k}) \big]. 
\end{eqnarray}
The derivatives in terms of $k_y$ and $k_z$ are obtained in a similar manner. 
Thus, with noting $N_{\rm u.c.} = N_{\rm site}/4$, 
the spectral weights for the multiorbital Hubbard model 
(\ref{eq:H_multiorbital}) are obtained in the forms 
\begin{eqnarray}
I_x &=& -\frac{1}{4N_{\rm site}} \Big[ 
\big\{ \langle (H_{\rm K}^{\rm nn})_{zx} \rangle
+ \langle (H_{\rm K}^{\rm nn})_{xy} \rangle \big\} 
\nonumber \\
&& \quad \quad 
+ 4 \big\{ \langle (H_{\rm K}^{\rm 3rd})_{zx} \rangle
+ \langle (H_{\rm K}^{\rm 3rd})_{xy} \rangle \big\} 
\Big], 
\label{eq:I_x Hub} 
\\
I_y &=& -\frac{1}{4N_{\rm site}} \Big[ 
\big\{ \langle (H_{\rm K}^{\rm nn})_{xy} \rangle
+ \langle (H_{\rm K}^{\rm nn})_{yz} \rangle \big\} 
\nonumber \\
&& \quad \quad 
+ 4 \big\{ \langle (H_{\rm K}^{\rm 3rd})_{xy} \rangle
+ \langle (H_{\rm K}^{\rm 3rd})_{yz} \rangle \big\} 
\Big], 
\label{eq:I_y Hub} 
\\
I_z &=& -\frac{1}{4N_{\rm site}} \Big[ 
\big\{ \langle (H_{\rm K}^{\rm nn})_{yz} \rangle
+ \langle (H_{\rm K}^{\rm nn})_{zx} \rangle \big\} 
\nonumber \\
&& \quad \quad 
+ 4 \big\{ \langle (H_{\rm K}^{\rm 3rd})_{yz} \rangle
+ \langle (H_{\rm K}^{\rm 3rd})_{zx} \rangle \big\} 
\Big], 
\label{eq:I_z Hub} 
\end{eqnarray}
where $(H_{\rm K}^{\rm nn})_{\zeta}$ and $(H_{\rm K}^{\rm 3rd})_{\zeta}$ are 
the nearest-neighbor and the third-neighbor matrix elements of 
the kinetic term in Eq.~(\ref{eq:H_multiorbital}) along the $\zeta$ chain, 
respectively. 

In the strong correlation limit, the kinetic energy corresponds to 
the spin and orbital exchange energy as discussed below. 
The expectation values of the kinetic terms 
in Eqs.~(\ref{eq:I_x Hub})-(\ref{eq:I_z Hub}) 
can be represented as coupling-constant derivatives;
\begin{eqnarray}
\langle (H_{\rm K}^{\rm nn})_{\zeta} \rangle &=& 
\Big\langle u_\zeta \frac{\partial}{\partial u_\zeta} 
H_{\rm Hub} \Big\rangle = 
\frac{\partial}{\partial \ln |u_\zeta|} 
\langle H_{\rm Hub} \rangle, 
\\
\langle (H_{\rm K}^{\rm 3rd})_{\zeta} \rangle &=& 
\frac{\partial}{\partial \ln |v_\zeta|} 
\langle H_{\rm Hub} \rangle. 
\end{eqnarray}
Here, we have used the Hellman-Feynman theorem to exchange 
the order of derivative and expectation value. 
In the strong correlation limit, we can replace the derivatives 
in terms of $u_\zeta$ and $v_\zeta$ of $\langle H_{\rm Hub} \rangle$
by the derivatives in terms of the exchange $J$ of 
the expectation values of the effective spin-orbital Hamiltonian 
(\ref{eq:H_SO}) as 
\begin{eqnarray}
\frac{\partial}{\partial \ln |u_\zeta|} \langle H_{\rm Hub} \rangle 
&=& 2 \frac{\partial}{\partial \ln |(J)_\zeta|} \langle H_{\rm SO} \rangle 
= 2 \langle (H_{\rm SO}^{\rm nn})_\zeta \rangle,
\label{eq:H_hub to H_SO 1}
\\
\frac{\partial}{\partial \ln |v_\zeta|} \langle H_{\rm Hub} \rangle 
&=& 2 \frac{\partial}{\partial \ln |(J_3)_\zeta|} \langle H_{\rm SO} \rangle
= 2 \langle (H_{\rm SO}^{\rm 3rd})_\zeta \rangle,
\label{eq:H_hub to H_SO 2}
\end{eqnarray}
where $(J)_\zeta = u_\zeta^2/U$ and $(J_3)_\zeta = v_\zeta^2/U$ are 
the nearest-neighbor and the third-neighbor exchange coupling constants 
along the $\zeta$ direction, respectively; and 
$(H_{\rm SO}^{\rm nn})_{\zeta}$ and $(H_{\rm SO}^{\rm 3rd})_{\zeta}$ are 
the nearest-neighbor and the third-neighbor matrix elements of 
Eqs.~(\ref{eq:H_nn}) and (\ref{eq:H_3rd}) along the $\zeta$ chain, 
respectively. 
In the model (\ref{eq:H_SO}), $(J)_\zeta = J = (t_\sigma^{\rm nn})/U$ and 
$(J_3)_\zeta = J_3 = (t_\sigma^{\rm 3rd})/U$, and they are 
independent of the direction $\zeta$. 
By substituting all the expectation values in 
Eqs.~(\ref{eq:I_x Hub})-(\ref{eq:I_z Hub}) 
by Eqs.~(\ref{eq:H_hub to H_SO 1}), (\ref{eq:H_hub to H_SO 2}), 
and the similar expressions, 
we obtain the formulas of the spectral weights for 
the effective spin-orbital coupled model in the strong correlation limit 
as given in Eqs.~(\ref{eq:I_x})-(\ref{eq:I_z}).




\begin{thebibliography}{99}
\bibitem{note3}
In the orbital ordered state below $T_{\rm O}$, 
there are two energy minima with equal energy; 
one is for the orbital occupations in Figs.~\ref{fig:orbital} (b)-(e) and
the other is for the occupations with exchanging $\bar{n}_1$ and $\bar{n}_2$. 
In the present MC calculations, one of two configurations 
is selected depending on the initial configuration, and 
the MC sampling is performed around the selected energy minimum, 
since the energy barrier between the two minima is exponentially large
in the first order transition.
\end{thebibliography}
\end{document}